\documentclass[aps,twocolumn,preprintnumbers,amsmath,amssymb,superscriptaddress,floatfix,nofootinbib,10pt]{revtex4}

\usepackage{graphicx,color,dcolumn,booktabs,bm}
\usepackage{epsfig}
\usepackage{epstopdf}
\usepackage{hyperref}
\usepackage{amsmath}
\allowdisplaybreaks[1]
\usepackage{amsfonts}
\usepackage{amssymb}
\usepackage{multirow}
\usepackage{makecell}
\usepackage[dvipsnames]{xcolor}
\usepackage{ulem}
\usepackage{longtable}
\usepackage{array}
\usepackage{subfigure}

\def\be{\begin{equation}}
\def\ee{\end{equation}}

\begin{document}

\title{Detecting the pure triangle singularity effect through the $\psi(2S) \to p \bar{p} \eta / p \bar{p} \pi^0$ process}
\author{Qi Huang}\email{huangqi@ucas.ac.cn}
\affiliation{University of Chinese Academy of Sciences (UCAS), Beijing 100049, China}
\author{Chao-Wei Shen} \email{shencw@ucas.ac.cn}
\affiliation{University of Chinese Academy of Sciences (UCAS), Beijing 100049, China}
\author{Jia-Jun Wu} \email{wujiajun@ucas.ac.cn}
\affiliation{University of Chinese Academy of Sciences (UCAS), Beijing 100049, China}

\date{\today}

\begin{abstract}

In this work, the triangle singularity mechanism is investigated in the $\psi(2S) \to p \bar{p} \eta / p \bar{p} \pi^0$ process.
The triangle loop composed by $J/\psi$, $\eta$ and $p$ has a singularity in the physical kinematic range for the $\psi(2S) \to p \bar{p} \eta / p \bar{p} \pi^0$ process,
and it would generate a very narrow peak in the invariant mass spectrum of $p\eta (\pi)$ around $1.56387$ GeV, which is far away from both the threshold and relative resonances.
In these processes, all the involved vertices are constrained by the experimental data. Thus, we can make a precise model independent prediction here.
It turns out that the peak in the $p\eta$ invariant mass spectrum is visible, while it is very small in the $p\pi^0$ invariant mass spectrum.
We expect this effect shown in $p \bar{p} \eta$ final state can be observed by the Beijing Spectrometer (BESIII) and Super Tau-Charm Facility (STCF) in the future.

\end{abstract}


\maketitle

\section{Introduction}\label{sec1}

The concept of triangle singularity is first proposed by L. D. Landau in 1959~\cite{Landau:1959fi}. In the following decades, it has been proved that it will play an important role in explaining many anomalous experimental observations.
For instance, the isospin breaking process $\eta(1405) \to \pi^0 f_0(980)$ observed by BESIII collaboration in 2012~\cite{BESIII:2012aa} was successfully explained in Refs.~\cite{Wu:2011yx,Aceti:2012dj,Wu:2012pg, Achasov:2015uua, Du:2019idk} by considering the triangle singularity produced in the $K \bar{K} K^\ast$ loop.
And in 2015, Ref.~\cite{Ketzer:2015tqa} explained the nature of $a_1(1420)$ with the $K^\ast \bar{K} K$ loop.
Especially in recent years, with the discovery of exotic states, such as $Z_c$~\cite{Ablikim:2013mio,Liu:2013dau,Xiao:2013iha,Ablikim:2013wzq,Ablikim:2013xfr,Ablikim:2013emm,Ablikim:2017oaf} and $P_c$ states~\cite{Aaij:2015tga,Aaij:2019vzc}, many researches on triangle singularity have been carried out~\cite{Wu:2011yx, Aceti:2012dj, Wu:2012pg, Ketzer:2015tqa, Wang:2013cya,Wang:2013hga, Achasov:2015uua, Liu:2015taa,Liu:2015fea,Guo:2015umn,Szczepaniak:2015eza, Guo:2016bkl, Bayar:2016ftu, Wang:2016dtb, Pilloni:2016obd, Xie:2016lvs, Szczepaniak:2015hya, Roca:2017bvy,
Debastiani:2017dlz, Samart:2017scf, Sakai:2017hpg, Pavao:2017kcr, Xie:2017mbe, Bayar:2017svj,Liang:2017ijf, Oset:2018zgc, Dai:2018hqb, Dai:2018rra, Guo:2019qcn, Liang:2019jtr, Nakamura:2019emd,Liu:2019dqc, Jing:2019cbw, Braaten:2019gfj, Sakai:2020ucu, Sakai:2020fjh, Molina:2020kyu, Braaten:2020iye, Alexeev:2020lvq, Ortega:2020ayw,Shen:2020gpw,Du:2019idk,Liu:2020orv,Achasov:2019wvw} (for a recent review, see Ref.~\cite{Guo:2019twa}).
Very recently, a new exotic state $X(2900)$ was observed by LHCb collaboration~\cite{Aaij:2020hon,Aaij:2020ypa}, and according to Ref.~\cite{Liu:2020orv}, it can also be related to a triangle singularity.

\begin{figure}[tbp]
	\centering
    \includegraphics[width=0.9\linewidth]{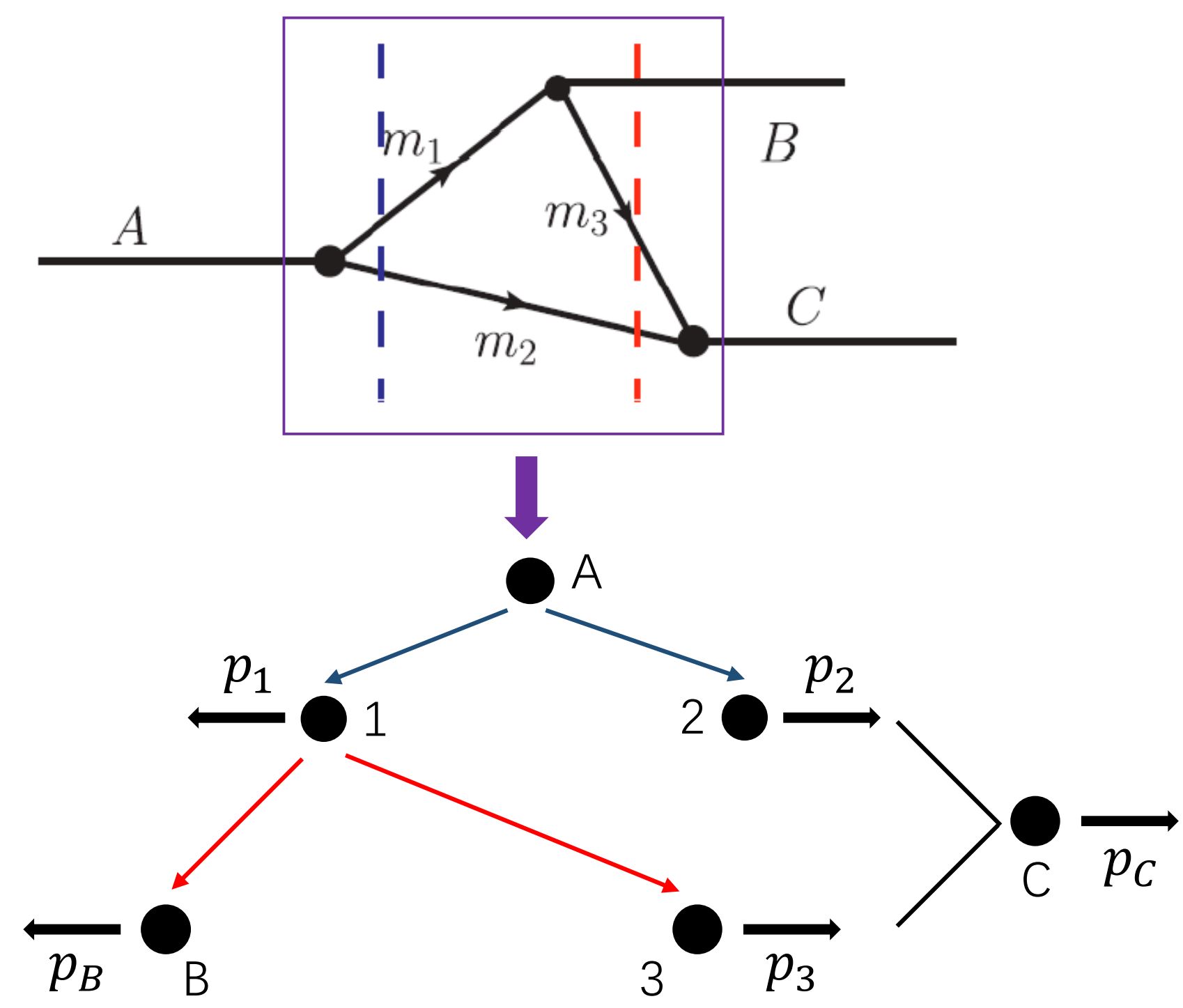}
    \caption{Kinematical mechanism of the production of a triangle singularity.}
    \label{fig:coleman-norton}
\end{figure}

The kinematics of triangle singularity can be briefly described by Fig.~\ref{fig:coleman-norton}.
In the figure, particle $A$ first decays into two particles 1 and 2.
In the center-of-mass frame of particle $A$, these two particles move in opposite directions.
If $m_A > m_1 + m_2$, this decay can really happen, which means that particles 1 and 2 can be classical particles.
Then particle 1 further decays into particle $B$ and particle 3, while particle 2 continues to move in its direction.
When particles 2 and 3 move in the same direction and the velocity of particle 3 is larger than that of particle 2, particle 3 will catch up with particle 2.
Eventually, particles 2 and 3 will actually collide and produce the final state C.
The whole process above is named as triangle singularity, which is the basic content of Coleman-Norton theorem~\cite{Shen:2020gpw}.

From Coleman-Norton theorem, we can easily see that the triangle singularity is a pure kinematical effect. Thus, such effect from the triangle loop is model independent and can be computed theoretically once all the vertices are known.
Obviously, the experimental detection of a pure triangle singularity must be interesting and important.
It will not only help us understand the triangle singularity itself, but also confirm the hadron loop mechanism.
Furthermore, it is useful to study the properties of hadrons, such as $X(3872)$~\cite{Guo:2019qcn}.

Unfortunately, although triangle singularities can be used to explain many experimental phenomena, the triangle singularity itself has not been fully confirmed by experiments for various reasons.
The first reason hindering the discovery of triangle singularity is the threshold effect.
For example, $Z_c(3900)$ and $Z_c(4025)$ are very close to the threshold of $\bar{D} D^\ast$ and $\bar{D}^\ast D^\ast$, respectively, and the $P_c$ state around 4.45 GeV is just near the $\chi_{c1}p$ threshold.
It is indicated in Ref.~\cite{Liu:2015taa} that when the threshold enhancement falls into the kinematic region of the triangle singularity, the distinctions between them would be very complicated.
Thus, a detectable pure triangle singularity shall be far away from the threshold enhancement.
In addition to the thresholds, the widths of internal particles should also be considered, i.e., the widths of internal particles should not be too wide. Otherwise, the shape of the triangle singularity will be a Breit-Wigner form, which might probably cause misunderstandings~\cite{Achasov:2015uua,Du:2019idk}.
Besides, if we want to determine a pure triangle singularity, a quantitative calculation should be necessary.
However, we found that among Refs.~\cite{Wu:2011yx, Aceti:2012dj, Wu:2012pg, Ketzer:2015tqa, Wang:2013cya,Wang:2013hga, Achasov:2015uua, Liu:2015taa,Liu:2015fea,Guo:2015umn,Szczepaniak:2015eza, Guo:2016bkl, Bayar:2016ftu, Wang:2016dtb, Pilloni:2016obd, Xie:2016lvs, Szczepaniak:2015hya, Roca:2017bvy,
Debastiani:2017dlz, Samart:2017scf, Sakai:2017hpg, Pavao:2017kcr, Xie:2017mbe, Bayar:2017svj,Liang:2017ijf, Oset:2018zgc, Dai:2018hqb, Dai:2018rra, Guo:2019qcn, Liang:2019jtr, Nakamura:2019emd,Liu:2019dqc, Jing:2019cbw, Braaten:2019gfj, Sakai:2020ucu, Sakai:2020fjh, Molina:2020kyu, Braaten:2020iye, Alexeev:2020lvq, Ortega:2020ayw,Shen:2020gpw,Du:2019idk,Liu:2020orv,Achasov:2019wvw}, most of them can only give the line shape since not all the involved vertices can be determined.
For example, although the exotic state $P_c(4450)$~\cite{Aaij:2015tga}, which is observed by LHCb collaboration in $\Lambda_b \to J/\psi p K$ process in 2015, can be interpreted as a triangle singularity effect caused by the $\chi_{c1} p \Lambda^\ast$ loop~\cite{Guo:2016bkl}, there is no experimental data to constraint the $\Lambda_b \chi_{c1} \Lambda^\ast$ vertex in this triangle loop diagram. Therefore, only a line shape can be presented theoretically.

In this work, we propose to detect a pure triangle singularity effect in the $p\eta/p\pi^0$ invariant mass spectrum in the $\psi(2S) \to p \bar{p} \eta / p \bar{p} \pi^0$ process with the triangle loop composed by $J/\psi$, $\eta$ and $p$.
This reaction can avoid the three aspects mentioned above to confirm the triangle singularity effect, the detailed explainations of these three points are as follows:
\begin{enumerate}
\item {\bf Far away from the threshold of the relative channel.}
By applying Coleman-Norton theorem, we get the position of triangle singularity in $m_{p\eta(\pi)}=1.56387$~GeV.
It is about 80~MeV away from the $p \eta$ threshold. It will certainly not mix with the $p\eta$ threshold effect as the width of the peak is quite small.
\item {\bf Narrow widths of all the intermediate particles in the loop.}
$J/\psi$, $\eta$ and $p$ all have a long life, so the peak of triangle singularity caused by such loop diagram must be very sharp, which means distinguishing it from $N^\ast$ is very easy.

\item {\bf The well known three vertices in the triangle loop.}
$\psi(2S)\to J/\psi \eta$, $J/\psi \to p \bar{p}$ and $p\eta \to p\eta /p \pi^0$ all can be constrained by the experimental data.
\end{enumerate}

Thus, we can make a very precise prediction on the significance of this triangle singularity theoretically.
We strongly recommend the experiments, especially BESIII and STCF (in the future), to do precise analysis on the $\psi(2S) \to p \bar{p} \eta / p \bar{p} \pi^0$ decay.

This paper is organized as follows.
After the introduction, a detailed calculation of $\psi(2S) \to p \bar{p} \eta / p \bar{p} \pi^0$ process via $J/\psi \eta p$ loop is given in Sec.~\ref{sec2}. Then the corresponding numerical results are shown in Sec.~\ref{sec3}, and finally a summary is presented.

\section{The triangle singularity caused by the $J/\psi \eta p$ loop}\label{sec2}

\subsection{The main mechanisms of the $\psi(2S) \to p \bar{p} \eta / p \bar{p} \pi^0$ process}\label{sec:2-mechanism}

\begin{figure}[htbp]
		\centering
    \includegraphics[width=0.45\linewidth]{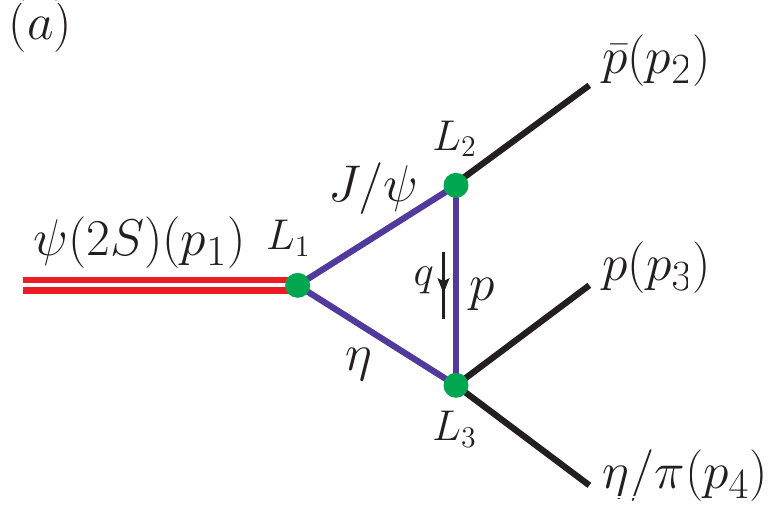} \
		\includegraphics[width=0.45\linewidth]{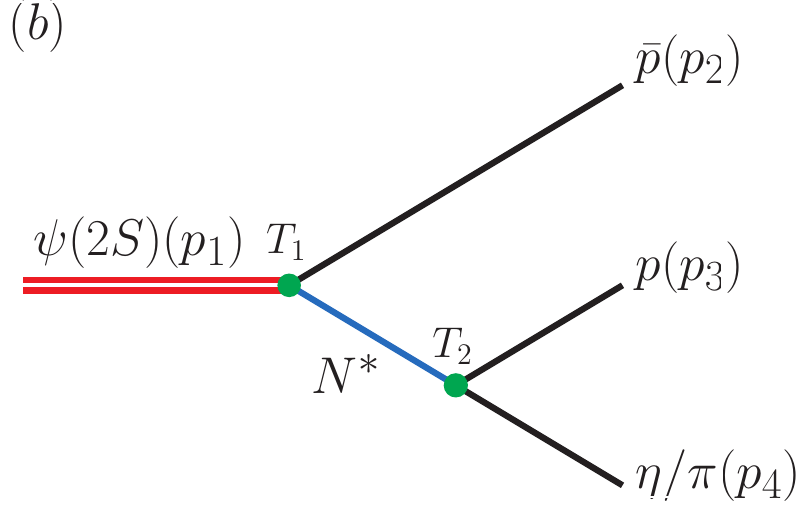}
  \caption{The Feynman diagrams describing the process $\psi(2S) \to p \bar{p} \eta / p \bar{p} \pi$. (a): loop diagram where triangle singularity happens; (b): tree diagram that called "background".}
  \label{fig:loop-and-background}
\end{figure}

The triangle loop diagram and tree diagram, which are shown in Fig.~\ref{fig:loop-and-background}, are both important for the $\psi(2S) \to p \bar{p} \eta / p \bar{p} \pi^0$ process. The triangle singularity we interested exists in the former, hence the latter is named as background in this work.
In the triangle loop diagram, $\psi(2S)$ first decays into $J/\psi$ and $\eta$, then $J/\psi$ decays into a $p\bar{p}$ pair and $\eta$ just moves along.
When $p$ catches up with $\eta$, they interact and re-scatter into the $p \eta~(p \pi)$ final state, and this case generates the triangle singularity.
In Fig.~\ref{fig:loop-and-background}(b), firstly $\psi(2S)$ decays into an anti-proton $\bar{p}$ and an excited nucleon $N^\ast$, then $N^\ast$ decays into the final state $p \eta~(p \pi)$.
It should be mentioned that in this work we ignore three diagrams, i.e., the loop diagram where the exchanged particle changes from $p$ to $\bar{p}$ and also exchanges the positions of the proton and anti-proton in the final states, the tree diagrams $\psi(2S) \to (\eta / \pi^0) (J/\psi \to \bar{p} p)$ and $\psi(2S) \to p \left(\bar{N}^\ast \to \bar{p} (\eta/\pi^0) \right)$.
The relevant reasons will be discussed in detail at the end of Sec. \ref{sec3A}.

The tree diagram is obviously dominant in Fig.~\ref{fig:loop-and-background}, thus, our main target is to check whether the narrow peak caused by the loop diagram is visible or not.
We adopt the effective Lagrangian approach to do the calculation and the general forms of the amplitudes that describing the Feynman diagrams in Fig.~\ref{fig:loop-and-background} are written as
\begin{eqnarray}
  \mathcal{M}^{\mathrm{Tree}}&=& \sum\limits_{N^\ast}\frac{ T_2 T_1}{m_{34}^2-m_{N^\ast}^2+i m_{N^\ast} \Gamma_{N^\ast}}, \label{eq:tree-general} \\
  \mathcal{M}^{\mathrm{Loop}} &=& i\int \frac{d^4 q}{(2\pi)^4}\frac{L_3 \mathcal{F}(p_3+p_4-q, m_\eta, \Lambda_\eta)} {(p_3+p_4-q)^2-m_\eta^2+im_\eta\Gamma_\eta}\nonumber\\
  &&\times \frac{L_1 \mathcal{F}(p_2+q,m_{J/\psi},\Lambda_{J/\psi})} {(p_2+q)^2-m_{J/\psi}^2+im_{J/\psi}\Gamma_{J/\psi}} \nonumber \\
  &&\times\frac{L_2 \mathcal{F} (q,m_p,\Lambda_p)} {q^2-m_p^2+i m_p \Gamma_p}, \label{eq:loop-general}
\end{eqnarray}
where $T_i$ and $L_i$ are the interactions of each vertices as given in Fig.~\ref{fig:loop-and-background}.
We introduce a form factor ${\mathcal{F}(q,m,\Lambda) = \frac{\Lambda^4}{(q^2-m^2)^2+\Lambda^4}}$ in Eq.~\eqref{eq:loop-general} to describe the structure effects of interaction vertices and off-shell effects of internal particles, and avoid the ultraviolet divergence.
When triangle singularity happens, all the exchanged particles are on-shell, so we have ${\mathcal{F}(q,m,\Lambda) = 1}$. It tells that the form factor will not affect the height of the peak caused by the triangle singularity.
In our calculation, for convenience, we set $\Lambda_{J/\psi,\eta,p} = m_{J/\psi,\eta,p} + \alpha \Lambda_{\text{QCD}}$, where $\alpha$ is a free parameter and $\Lambda_{\text{QCD}}=0.22$~GeV.

Then, all the invariant mass spectrums we need can be obtained by
\begin{eqnarray}
  d\Gamma = \frac{\sum \left|\mathcal{M}^{\mathrm{Tree}}+\mathcal{M}^{\mathrm{Loop}}\right|^2}{96(2\pi)^3 m_{\psi(2S)}^3} dm_{23}^2 dm_{34}^2,
\end{eqnarray}
where $\sum$ denotes the summations over spins of related particles, $m^2_{23}=(p_2+p_3)^2$, and $m^2_{34}=(p_3+p_4)^2$.

\subsection{Effective Lagrangians depicting the interactions of the three-particle vertices}\label{sec:2-Lagrangian}

Here we introduce the effective Lagrangians describing the interactions of all the vertices in Fig.~\ref{fig:loop-and-background}.
For the interaction between $\psi(2S)$, $J/\psi$ and $\eta$, we adopt the general coupling form of vector ($\mathcal{V}$) and pesudo-scalar ($\mathcal{P}$) mesons
\begin{eqnarray}
  \mathcal{L}_{\mathcal{V} \mathcal{V} \mathcal{P}}= g_{\mathcal{V} \mathcal{V} \mathcal{P}}\varepsilon^{\mu\nu\alpha\beta} \partial_\mu \mathcal{V}_\nu \partial_\alpha \mathcal{V}_\beta \mathcal{P}.\label{eq:Lagrangian-VVP}
\end{eqnarray}
For the interactions between vector meson and nucleon states, i.e. $\mathcal{V} N^\ast N$, the effective Lagrangians are~\cite{Tsushima:1996xc,Tsushima:1998jz,Zou:2002yy,Ouyang:2009kv,Wu:2009md,Cao:2010km,Cao:2010ji}
\begin{eqnarray}
  \mathcal{L}_{\mathcal{V} N \bar{N}}&=&-g_{\mathcal{V} N \bar{N}} \bar{N} \gamma_\mu \mathcal{V}^\mu N,\label{eq:Lagrangian-N}\\
  \mathcal{L}_{\mathcal{V} P_{11} \bar{N}} &=&-g_{\mathcal{V} P_{11} \bar{N}} \bar{N} \gamma_\mu \mathcal{V}^\mu P_{11} + h.c.,\label{eq:Lagrangian-P11-V}\\
  \mathcal{L}_{\mathcal{V} S_{11} \bar{N}} &=& -g_{\mathcal{V} S_{11} \bar{N}} \bar{N} \gamma_5 \gamma_\mu \mathcal{V}^\mu S_{11} + h.c.,\label{eq:Lagrangian-S11-V}\\
  \mathcal{L}_{\mathcal{V} D_{13} \bar{N}} &=& -g_{\mathcal{V} D_{13} \bar{N}} \bar{N} \mathcal{V}_\mu D_{13}^\mu + h.c.,\label{eq:Lagrangian-D13-V}
\end{eqnarray}
where $P_{11}$, $S_{11}$, $D_{13}$ represent the fields of excited nucleons with quantum numbers $J^P = 1/2^+$, $1/2^-$ and $3/2^-$, respectively.
Finally, the interactions between $N^\ast$, $N$ and pesudo-scalar meson $\mathcal{P}$ can be described by the effective Lagrangians as follows~\cite{Xu:2015qqa},
\begin{eqnarray}
  \mathcal{L}_{\mathcal{P} N P_{11}}&=&-\frac{g_{\mathcal{P} N P_{11}}}{2m_N} \bar{N} \gamma_5 \gamma_\mu \partial^\mu\mathcal{P} P_{11} + h.c.,\label{eq:Lagrangian-P11-P}\\
  \mathcal{L}_{\mathcal{P} N S_{11}}&=&-g_{\mathcal{P} N S_{11}} \bar{N} \mathcal{P} S_{11} + h.c.,\label{eq:Lagrangian-S11-P}\\
  \mathcal{L}_{\mathcal{P} N D_{13}}&=&-\frac{g_{\mathcal{P} N D_{13}}}{m_N^2} \bar{N} \gamma_5 \gamma^\mu \partial_\mu\partial_\nu\mathcal{P} D_{13}^\nu + h.c.\label{eq:Lagrangian-D13-P},
\end{eqnarray}
with $\mathcal{P}$ being the pseudoscalar octet, i.e.
\begin{eqnarray}
\mathcal{P} =
 \left(
 \begin{array}{ccc}
\sqrt{\frac{1}{2}} \pi^{0}+ \sqrt{\frac{1}{6}} \eta & \pi^{+} & K^{+}\\
\pi^{-} & -\sqrt{\frac{1}{2}} \pi^{0}+ \sqrt{\frac{1}{6}} \eta &  K^{0}\\
 K^{-} & \bar{K}^{0} & -\sqrt{\frac{2}{3}} \eta
 \end{array}
 \right).
\end{eqnarray}

Therefore, the interaction between $\psi(2S)$ and $N^\ast \bar{p}$, i.e. $T_1$, can be extracted from Eqs.~\eqref{eq:Lagrangian-P11-V}-\eqref{eq:Lagrangian-D13-V}.
Taking the $N^\ast$ being $N(1535)$ as an example, we use Eq.~\eqref{eq:Lagrangian-S11-V} and get
\begin{eqnarray}
  T_1 = -g_{\psi(2S) \bar{p} N(1535)} \bar{u}(p_{34}) \gamma_5 \gamma_\mu \epsilon_{\psi(2S)}^\mu v(p_2),
\end{eqnarray}
where $u$, $v$ are the spin wave functions of the particle and anti-particle, respectively, and the notation is taken as $\sum\limits_{\mathrm{spin}} u\bar{u}=p\!\!\!/+m$, also, $\epsilon$ indicates the polarization vector of the corresponding particle, and $p_{34}=p_3+p_4$.

Similarly, the coupling between $N^\ast$ and $p\eta/p\pi^0$, can be got from Eqs.~\eqref{eq:Lagrangian-P11-P}-\eqref{eq:Lagrangian-D13-P}.
We also take $N(1535)$ as an example, and by using Eq.~\eqref{eq:Lagrangian-S11-P} we have
\begin{eqnarray}
  T_2 = -g_{N(1535) p \mathcal{P}} \bar{u}(p_3) u(p_{34}).
\end{eqnarray}

At last, $L_1$ and $L_2$ can be extracted from Eq.~\eqref{eq:Lagrangian-VVP} and Eq.~\eqref{eq:Lagrangian-N} as
\begin{eqnarray}
  L_1 &=& g_{\psi(2S) \psi p} \varepsilon^{\mu\nu\alpha\beta} p_{1\mu} \epsilon_{\psi(2S)\nu} (p_{2}+q)_\alpha \epsilon^\ast_{J/\psi\beta},\\
  L_2 &=& -g_{\psi p \bar{p}} \bar{u}(q) \gamma_\mu \epsilon_{J/\psi}^\mu v(p_2).
\end{eqnarray}

\subsection{Describing $p\eta \to p\eta$ and $p\eta \to p \pi^0$ processes}\label{sec:2-L3}

We have written $L_1$, $L_2$, $T_1$ and $T_2$ in Sec.~\ref{sec:2-Lagrangian}, while the specific expression of $L_3$ in Eq.~\eqref{eq:loop-general} is still unknown.
In this subsection we will present how to get $L_3$ in detail.

For the $p\bar{p}\eta$ case, $L_3$ represents the elastic scattering of $p\eta$.
The main contributions should come from the s-channel contributions of excited nucleons $N^\ast$.
Since the position of the triangle singularity is 1.56387~GeV, we find that only $N(1535)$ and $N(1650)$ have considerable decay rates to $p\eta$ in RPP~\cite{Zyla:2020zbs} around this energy.
Therefore, we can use effective Lagrangians given in Sec.~\ref{sec:2-Lagrangian} to write $L_3$ as
\begin{eqnarray}
  L_3&=&g_{N(1535)p\eta}^2  \frac{\bar{u}(p_3) \left(p\!\!\!/_{34}+m_{N(1535)} \right) u(q)}{m_{34}^2 - m_{N(1535)}^2 + i m_{N(1535)} \Gamma_{N(1535)}}\nonumber\\
  &&+g_{N(1650)p\eta}^2  \frac{\bar{u}(p_3) \left(p\!\!\!/_{34}+m_{N(1650)} \right) u(q)}{m_{34}^2 - m_{N(1650)}^2 + i m_{N(1650)} \Gamma_{N(1650)}}.\nonumber\\ \label{eq:L3-eta}
\end{eqnarray}
Here, the phase angle between these two $N^\ast$ states is ignored. Actually, from the results shown in Sec.~\ref{sec:2-Amp}, we find that the contribution of $N(1650)$ is not important.

\begin{figure}[htbp]
	\centering
  \includegraphics[width=0.9\linewidth]{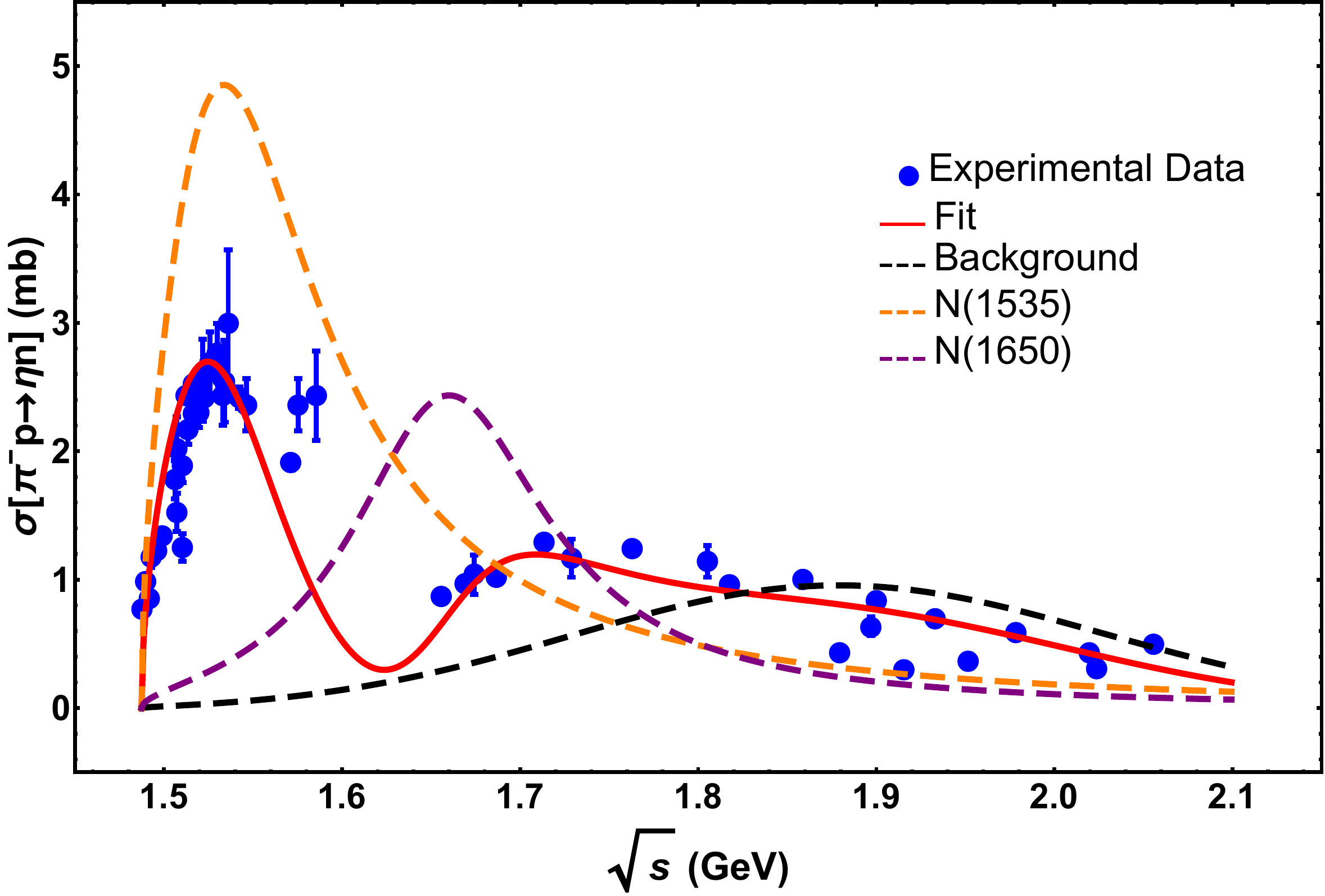}
  \caption{The fitted results of the cross section of $\pi^- p \to \eta n$. The blue points are the experimental data given in Refs.~\cite{Prakhov:2005qb,Deinet:1969cd,Richards:1970cy,Brown:1979ii,Crouch:1980vw,Feltesse:1975nz}, the red solid line is our fitted result, and the black, orange and purple dashed lines are the contributions of the background, $N(1535)$ and N(1650), respectively.}
  \label{fig:cross-section-fit}
\end{figure}

For the $p\bar{p}\pi^0$ case, $L_3$ represents the $\eta p \to \pi^0 p$ scattering, we can extract it from the experimental data of $\pi^- p \to \eta n$. Firstly, we fit the experimental data of $\pi^- p(q) \to \eta(p_4) n(p_3)$ process~\cite{Prakhov:2005qb,Deinet:1969cd,Richards:1970cy,Brown:1979ii,Crouch:1980vw,Feltesse:1975nz}
with the following amplitude
\begin{eqnarray}
  \mathcal{M}_{\pi^- p \to \eta n} &=& g_B \bar{u}(q) u(p_3)e^{-a_B(\sqrt{(p_{3}+p_4)^2}-\Sigma_B)^2}\nonumber\\
  &&+ \sum\limits_ke^{i \phi_{N^\ast_k}} \frac{g_{N^\ast_k} \bar{u}(q) (p\!\!\!/_{3}+p\!\!\!/_{4}+m_{N^\ast_k})u(p_3)}{(p_{3}+p_4)^2-m_{N^\ast_k}^2+im_{N^\ast_k}\Gamma_{N^\ast_k}},\nonumber\\
  \label{eq:fit}
\end{eqnarray}
where the first term represents the background, and the second term is the contributions of various $N^\ast$ states exchange.
During the fit, we find that only the two $N^\ast$ states, i.e., $N(1535)$ and $N(1650)$, are needed except for the background.
The parameters we get by fitting the cross section of $\pi^- p \to \eta n$ are given in Table~\ref{tab:cross-section-fit} and the fitted results are presented in Fig.~\ref{fig:cross-section-fit}.
By using isospin relation naively, we can finally get
\begin{eqnarray}
  L_3=\frac{1}{\sqrt{2}}\mathcal{M}_{\pi^- p \to \eta n}\label{eq:L3-pi}.
\end{eqnarray}

\begin{table}[htpb]
	\renewcommand\arraystretch{1.5}
  \centering
	\caption{The parameters we get by fitting the cross section of $\pi^- p \to \eta n$ given in Ref.~\cite{Prakhov:2005qb,Deinet:1969cd,Richards:1970cy,Brown:1979ii,Crouch:1980vw,Feltesse:1975nz}. We should emphasize here that in this table $g_{N(1535)} = g_{N(1535)\pi^- p} \times g_{N(1535)\eta p}$ and $g_{N(1650)} = g_{N(1650)\pi^- p} \times g_{N(1650)\eta p}$, and $g_{N(1535)\pi^- p}$, $g_{N(1535)\eta p}$, $g_{N(1650)\pi^- p}$, $g_{N(1650)\eta p}$ are extracted from the branching ratios of related decays given in RPP \cite{Zyla:2020zbs}.}
	\label{tab:cross-section-fit}
  \begin{tabular}{ccc}
  \toprule[1pt]
  Coupling constants &$\quad$& Values\\
  \midrule[1pt]
  $g_B$ &$\quad$& 12.992$\pm$0.102 GeV$^{-1}$\\
  $a_B$ &$\quad$& 10.814$\pm$0.419 GeV$^{-2}$\\
  $\Sigma_B$ &$\quad$& 1.873$\pm$0.003 GeV\\
  $\phi_{N(1535)}$ &$\quad$& 0.00841$\pm$0.001\\
  $g_{N(1535)}$ &$\quad$& 2.80$\pm$0.72\\
  $\phi_{N(1650)}$ &$\quad$& 0.921$\pm$0.018\\
  $g_{N(1650)}$ &$\quad$& 1.525$\pm$ 0.583\\
  \bottomrule[1pt]
  \end{tabular}
\end{table}

\subsection{Amplitudes of $\psi(2S) \to p \bar{p} \eta / p \bar{p} \pi^0$ process}\label{sec:2-Amp}

After all the preparations made above, we can write the amplitudes of the $\psi(2S) \to p \bar{p} \eta / p \bar{p} \pi^0$ process.

For the $\psi(2S) \to p \bar{p} \eta$ process, one can see that  for $\psi(2S)$ the main contribution to $p\bar{p}\eta$ channel is $\psi(2S) \to \bar{p} N(1535) \to \bar{p} p \eta$~\cite{Zyla:2020zbs}.
Thus, at tree diagram level, we only need to consider the contribution from $N(1535)$ and have
\begin{eqnarray}
  \mathcal{M}^{\mathrm{Tree}} &=& (g_{\psi(2S)\bar{p}N(1535)} \times g_{N(1535)p\eta}) \epsilon_{\psi(2S)}^\mu \nonumber\\
  &&\times\frac{ \bar{u}(p_3) \left(p\!\!\!/_{34}+m_{N(1535)} \right) \gamma_5 \gamma_\mu  v(p_2)}{m_{34}^2 - m_{N(1535)}^2 + i m_{N(1535)} \Gamma_{N(1535)}},\nonumber\\
\end{eqnarray}
and it is assumed to be dominant. We extract the coupling constants $(g_{\psi(2S)\bar{p}N(1535)} \times g_{N(1535)p\eta}) = 0.0013769 \pm 0.0000673$ from the branching ratio $\mathcal{B}(\psi(2S) \to \bar{p}N(1535) \to \bar{p} p \eta) = (2.20 \pm 0.35) \times 10^{-5}$.
And using Eq.~\eqref{eq:L3-eta}, we have the amplitude of the loop diagram
\begin{eqnarray}
  \mathcal{M}^{\mathrm{Loop}} &=&i\int \frac{d^4 q}{(2\pi)^4}  g_{\psi(2S) \psi \eta} g_{J/\psi p \bar{p}} \varepsilon^{\mu \nu \alpha \beta} p_{1\mu} \epsilon_{\psi(2S)\nu} k_{1\alpha} \nonumber\\
  &&\times \bar{u}(p_3) \left( \frac{g_{N(1535)p\eta}^2 \left(p\!\!\!/_{34}+m_{N(1535)} \right)}{m_{34}^2 - m_{N(1535)}^2 + i m_{N(1535)} \Gamma_{N(1535)}} \right.\nonumber\\
  &&\left. + \frac{g_{N(1650)p\eta}^2 \left(p\!\!\!/_{34}+m_{N(1650)} \right)}{m_{34}^2 - m_{N(1650)}^2 + i m_{N(1650)} \Gamma_{N(1650)}} \right)\nonumber\\
  && \times (q\!\!\!/+m_q) \gamma_\beta v(p_2)\frac{\mathcal{F}(p_3+p_4-q, m_\eta, \Lambda_\eta)} {(p_3+p_4-q)^2-m_\eta^2+im_\eta\Gamma_\eta}\nonumber \\
  &&\times\frac{\mathcal{F}(p_2+q,m_{J/\psi},\Lambda_{J/\psi})} {(p_2+q)^2-m_{J/\psi}^2+im_{J/\psi}\Gamma_{J/\psi}}\nonumber\\
  &&\times\frac{\mathcal{F} (q,m_p,\Lambda_p)} {q^2-m_p^2+i m_p \Gamma_p}.\label{eq:p-p-eta-loop}
\end{eqnarray}
All the involved coupling constants are extracted from relevant branching ratios~\cite{Zyla:2020zbs} and the values are listed in Table~\ref{tab:eta-coupling}.
\begin{table}[htpb]
	\renewcommand\arraystretch{1.5}
  \centering
	\caption{The values of the coupling constants involved in Eq.~\eqref{eq:p-p-eta-loop}.}
	\label{tab:eta-coupling}
  \begin{tabular}{ccc}
  \toprule[1pt]
  Coupling constant & Branching ratio& Value\\
  \midrule[1pt]
  $g_{\psi(2S)\psi\eta}$ &$(3.37 \pm 0.05) \times 10^{-2}$& 0.218 $\pm$ 0.003\\
  $g_{J/\psi p \bar{p}}$ &$(2.121 \pm 0.029) \times 10^{-3}$& 0.0016 $\pm$ 0.0002\\
  $g_{N(1535)p\eta}$ &$30\% - 51\%$& 2.59 $\pm$ 0.62\\
  $g_{N(1650)p\eta}$ &$11\% - 31\%$& 1.24 $\pm$ 0.31\\
  \bottomrule[1pt]
  \end{tabular}
\end{table}

We apply the similar method to the $\psi(2S) \to p \bar{p} \pi^0$ process. In this case, $N(1535)$, $N(1650)$, $N(1440)$ and $N(1520)$ will give contributions to $p \bar{p} \pi^0$ channel~\cite{Zyla:2020zbs}. Thus, the tree-level amplitude can be written as
\begin{eqnarray}
  \mathcal{M}^{\mathrm{Tree}} &=& (g_{\psi(2S)\bar{p}N(1535)} \times g_{N(1535)p\pi^0}) \epsilon_{\psi(2S)}^\mu \nonumber\\
  &&\times\frac{ \bar{u}(p_3) \left(p\!\!\!/_{34}+m_{N(1535)} \right) \gamma_5 \gamma_\mu  v(p_2)}{m_{34}^2 - m_{N(1535)}^2 + i m_{N(1535)} \Gamma_{N(1535)}}\nonumber\\
  && + (g_{\psi(2S)\bar{p}N(1650)} \times g_{N(1650)p\pi^0}) \epsilon_{\psi(2S)}^\mu \nonumber\\
  &&\times\frac{ \bar{u}(p_3) \left(p\!\!\!/_{34}+m_{N(1650)} \right) \gamma_5 \gamma_\mu  v(p_2)}{m_{34}^2 - m_{N(1650)}^2 + i m_{N(1650)} \Gamma_{N(1650)}}\nonumber\\
  &&+ i \frac{(g_{\psi(2S)\bar{p}N(1440)} \times g_{N(1440)p\pi^0})}{2m_{N(1440)}} \epsilon_{\psi(2S)}^\mu \nonumber\\
  &&\times\frac{ \bar{u}(p_3) \gamma_5 p\!\!\!/_4 \left(p\!\!\!/_{34}+m_{N(1440)} \right) \gamma_\mu  v(p_2)}{m_{34}^2 - m_{N(1440)}^2 + i m_{N(1440)} \Gamma_{N(1440)}}\nonumber\\
  &&- \frac{(g_{\psi(2S)\bar{p}N(1520)} \times g_{N(1520)p\pi^0})}{m_{N(1520)}^2} \epsilon_{\psi(2S)}^\mu \nonumber\\
  &&\times\frac{ \bar{u}(p_3) \gamma_5 p\!\!\!/_4 \left(p\!\!\!/_{34}+m_{N(1520)} \right) \tilde{G}_{\mu\nu} p_4^\nu  v(p_2)}{m_{34}^2 - m_{N(1520)}^2 + i m_{N(1520)} \Gamma_{N(1520)}},\nonumber\\\label{eq:p-p-pi-tree}
\end{eqnarray}
where $\tilde{G}_{\mu\nu} = -g_{\mu\nu} + \frac{\gamma_\mu \gamma_\nu}{3} + \frac{\gamma_\mu (p_{3\nu}+p_{4\nu}) - \gamma_\nu (p_{3\mu}+p_{4\mu})}{3m_{N(1520)}} + \frac{2 (p_{3\mu}+p_{4\mu}) (p_{3\nu}+p_{4\nu})}{3m_{N(1520)}^2}$~\cite{Xu:2015qqa}, and the relevant coupling constants are listed in Table~\ref{tab:p-p-pi-tree}.
It should be noticed that here all the phase angles between any $N^\ast$ are also ignored, and we will prove that this treatment does not conflict with the present experimental results later.
Using Eqs.~\eqref{eq:fit}-\eqref{eq:L3-pi}, the amplitude of loop diagram writes
\begin{eqnarray}
  \mathcal{M}^{\mathrm{Loop}} &=&i\int \frac{d^4 q}{(2\pi)^4} g_{\psi(2S)\psi\eta} g_{J/\psi p \bar{p}} \varepsilon^{\mu\nu\alpha\beta} p_{1\mu} \epsilon_{\psi(2S)\nu} k_{1\alpha} \nonumber\\
  &&\times \bar{u}(p_3) \left( \frac{g_B}{\sqrt{2}} e^{-a_B(\sqrt{p_{34}^2}-\Sigma_B)^2} +\sum\limits_k \frac{e^{i \phi_{N^\ast_k}}}{\sqrt{2}}\right. \nonumber\\
  &&\times\left. \frac{g_{N^\ast_k}}{p_{34}^2-m_{N^\ast_k}^2+im_{N^\ast_k}\Gamma_{N^\ast_k}} (p\!\!\!/_{34} +m_{N^\ast_k}) \right)\nonumber\\
  &&\times(q\!\!\!/+m_p) \gamma_\beta v(p_2)\nonumber\\
  &&\times\frac{\mathcal{F}(p_3+p_4-q, m_\eta, \Lambda_\eta)} {(p_3+p_4-q)^2-m_\eta^2+im_\eta\Gamma_\eta}\nonumber\\
  &&\times \frac{\mathcal{F}(p_2+q,m_{J/\psi},\Lambda_{J/\psi})} {(p_2+q)^2-m_{J/\psi}^2+im_{J/\psi}\Gamma_{J/\psi}}\nonumber\\
  &&\times\frac{\mathcal{F} (q,m_p,\Lambda_p)} {q^2-m_p^2+i m_p \Gamma_p}.
\end{eqnarray}

\begin{table}[htpb]
	\renewcommand\arraystretch{1.5}
  \centering
	\caption{The values of the coupling constants involved in Eq.~\eqref{eq:p-p-pi-tree}.}
	\label{tab:p-p-pi-tree}
  \begin{tabular}{ccc}
  \toprule[1pt]
  Intermediate $N^\ast$ \ &  Branching ratio~\cite{Zyla:2020zbs} \ & Coupling constants\\
    &  &$g_{\psi(2S)\bar{p}N^*} \times g_{N^*p\pi^0}$\\
 \midrule[1pt]
  $N(1440) $ & $(7.3^{+1.7}_{-1.5})\times 10^{-5}$  & $(2.39^{+0.10}_{-0.09})\times 10^{-3}$\\
  $N(1520) $ & $(6.4^{+2.3}_{-1.8})\times 10^{-6}$  & $(1.62^{+0.23}_{-0.21})\times 10^{-3}$\\
  $N(1535) $ & $(2.5\pm 1.0)\times 10^{-5}$  & $(7.48 \pm 0.401)\times 10^{-4}$\\
  $N(1650) $ & $(3.8^{+1.4}_{-1.7})\times 10^{-5}$  & $(9.44^{+0.81}_{-0.83})\times 10^{-4}$\\
  \bottomrule[1pt]
  \end{tabular}
\end{table}

\section{Numerical results}\label{sec3}

\subsection{Triangle singularity in $\psi(2S) \to p \bar{p} \eta$ process}
\label{sec3A}

\begin{figure}[htbp]
	\centering
  \includegraphics[width=0.95\linewidth]{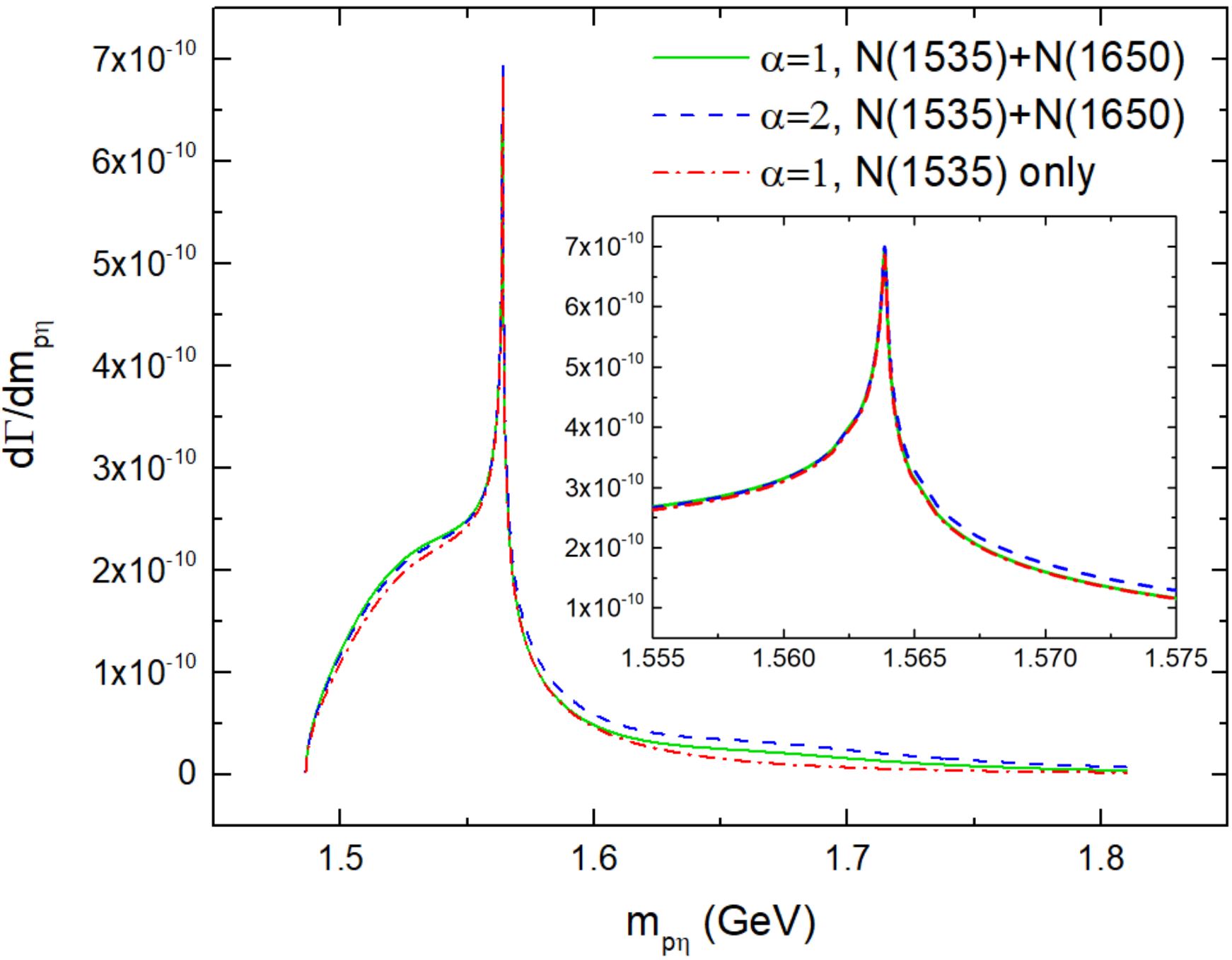}
  \caption{The $p\eta$ invariant mass spectrum of $\mathcal{M}^{\mathrm{Loop}}$. The green solid line represents that $\alpha$ in the formfactor is 1 and both $N(1535)$ and $N(1650)$ are included. The blue dashed line represents that $\alpha$ is 2 and both $N(1535)$ and $N(1650)$ are included. The red dash-dot line represents that $\alpha$ is 1 and only $N(1535)$ is considered.}
  \label{fig:p-p-eta-alpha}
\end{figure}

After all the preparations in Sec.~\ref{sec2}, we can calculate these two processes and their numerical results are given.
First of all, we need to prove our previous arguments that the form factor $\mathcal{F}(q,m,\Lambda)$ and $N(1650)$ in $\bar{p}p\eta$ channel will not affect the height of triangle singularity too much.
The $m_{p\eta}$ invariant mass spectrum of $\mathcal{M}^{\mathrm{Loop}}$ is given in Fig.~\ref{fig:p-p-eta-alpha}.
In Fig.~\ref{fig:p-p-eta-alpha}, we consider different values of the parameter $\alpha$ in the form factor, and the cases whether $N(1650)$ is included.
It is clear that the peak caused by triangle singularity is very sharp and its width is only about 1~MeV.
In addition, we can see that the peak around where triangle singularity happens does change little when $\alpha$ changes from 1 to 2, and the contribution of $N(1650)$ is negligible.

\begin{figure}[htbp]
	\centering
  \includegraphics[width=0.9\linewidth]{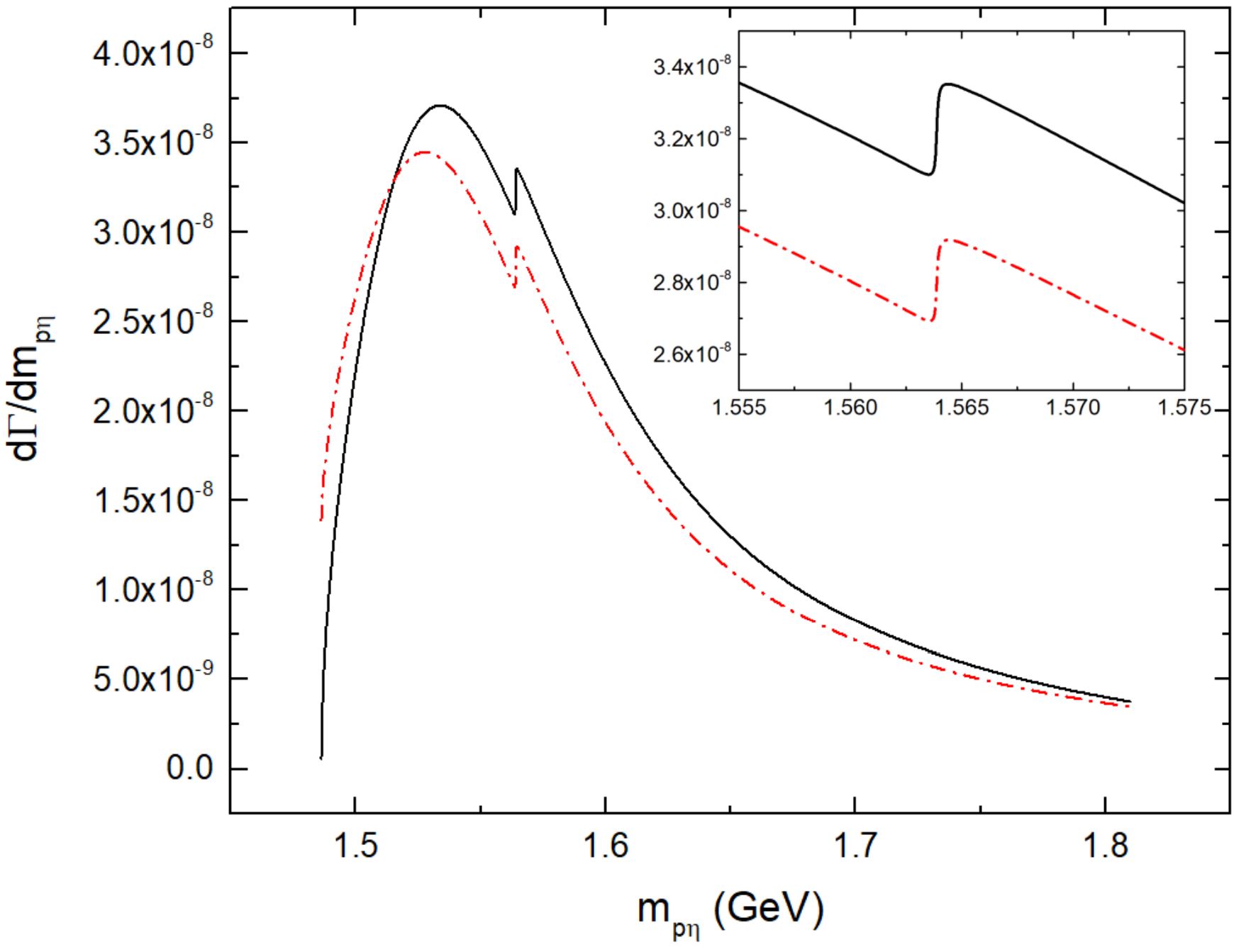}
  \caption{The $p\eta$ invariant mass distributions of the $\psi(2S) \to p \bar{p} \eta$ process. For black solid line,  we consider the interface between $\mathcal{M}^{\mathrm{Loop}}$ and $\mathcal{M}^{\mathrm{Tree}}$ with the relative phase angle being 0, and here both $N(1535)$ and $N(1650)$ exchange are included and $\alpha$ is set to 1. For the red dashed-dotted line, there includes one more tree diagram $\psi(2S) \to \eta (J/\psi \to \bar{p} p)$ with the relative phase angle being 0 and having a $m_{p\bar{p}}<3.067$~GeV cut following Ref.~\cite{Ablikim:2013vtm}.}
  \label{fig:p-p-eta}
\end{figure}

Then we consider the interference between $\mathcal{M}^{\mathrm{Loop}}$ and $\mathcal{M}^{\mathrm{Tree}}$ to see whether this peak is visible in the experiments or not.
The $p\eta$ invariant mass distribution of the $\psi(2S) \to p \bar{p} \eta$ process is given by the black solid line in Fig.~\ref{fig:p-p-eta}.
The Breit-Wigner shape of $N(1535)$ from the tree diagram is quite clear and we can see that there exists a visible enhancement, which is caused purely by triangle singularity, on the right shoulder of this peak.
We also amplify the enhancement part and draw it in detail in Fig.~\ref{fig:p-p-eta}. The width of this structure is about 5~MeV.
According to our calculation, the enhancement of the peak comparing to the background is about $10\%$. It implies that if we have 4 billion $\psi(2S)$ events, the number of events from this peak is about 120. During the discussions~\cite{Discussion:Lyu}, the experimentalists tell us that if the resolution of experiments can reach 2-3~MeV, this structure will be visible experimentally.
However, the highest resolution of BESIII detector is around 4.3~MeV at present, then such enhancement will be absorbed and is hard to distinguish.
Therefore, we expect the future experiments such as STCF could have higher resolution.

We have to emphasize that the prediction of $10\%$ enhancement above is based on the assumption that the phase angle in the interference between the tree and loop diagrams is 0.
Although it is the only assumption in our calculation, once this phase angle has a specific value, this enhancement might become invisible.

Since the vertex $L_3$ in the $p\bar{p}\eta$-loop case represents the elastic scattering $p\eta \to p\eta$, according to the Schmid theorem~\cite{Schmid:1967ojm,Debastiani:2018xoi}, the triangle singularity from Fig.~\ref{fig:loop-and-background} ($a$) will become negligible after considering the $\psi(2S) \to \eta (J/\psi \to \bar{p} p)$ process of tree diagram.
However, the realistic $\psi(2S) \to \bar{p} p\eta$ reaction is too complicated to apply the Schmid theorem directly. At least, the contribution of triangle singularity is still visible for the following three reasons.

Firstly, Schmid theorem tells that if the $p\eta \to p\eta$ process, i.e., the $L_3$ in Fig.~\ref{fig:loop-and-background} (a), is purely elastic, then the triangle singularity caused by the loop diagram will be negligible.
This is because the contribution of the loop diagram only affects the tree diagram $\psi(2S)\to \eta(J/\psi\to p\bar p)$ though a phase factor $\mbox{exp}(2i\delta^{p\eta}_0)$.
Nevertheless, we should emphasize here that the $p\eta \to p\eta$ process can not be recognized as a purely elastic process, because another channel $p \pi$ definitely couples with it.
In other words, the pure scattering amplitude of $S$-matrix $S_{p\eta \to p\eta}$ can not satisfy the unitary by itself, that is, $|S_{p\eta \to p\eta}|^2<1$.
At least we need to include the $p\pi$ channel to extend it to a $2\times 2$ matrix to have $|S|^2=I$.
And in our calculation, $p\eta \to p\eta$ process is mainly described by the s-channel $N^\ast$ states exchange, where the $p\pi$ contribution has already been included effectively in the imaginary parts of the $N^\ast$ propagators.
There are similar discussions in Ref.~\cite{Debastiani:2018xoi}.
As a result, we can not directly apply the Schmid theorem here.

Secondly, in the $\psi(2S) \to \bar{p}p\eta$ reaction, the tree diagram process $\psi(2S) \to \bar{p} (N^\ast \to p\eta)$ definitely plays an important role.
Compared to the conclusion of the Schmid theorem, which is
$|t_{J/\psi}^{\mathrm{tree}}+t_{\mathrm{elastic}}^{\mathrm{loop}}|^2 =
|t_{J/\psi}^{\mathrm{tree}} e^{i\delta} |^2 = |t_{J/\psi}^{\mathrm{tree}}|^2$,
the additional term of tree diagram exchanging $N^\ast$, $t^{\mathrm{tree}}_{N^\ast}$, will modify the total amplitude as
$|t_{J/\psi}^{\mathrm{tree}}+t^{\mathrm{loop}}
+t^{\mathrm{tree}}_{N^\ast}|^2$, where $t^{\mathrm{loop}}$ includes more than $t_{\mathrm{elastic}}^{\mathrm{loop}}$ as discussed above.
Thus, some terms must exist to reflect the interferences between $t^{\mathrm{loop}}$ and $t^{\mathrm{tree}}_{N^\ast}$, and it would lead to a weak signal of triangle singularity in the $p\eta$ invariant mass spectrum as shown in Fig.~\ref{fig:p-p-eta}.

The third and the most important reason is that when analyzing the experimental data of the $\psi(2S) \to p\bar{p}\eta$ process, a cut $m_{p\bar{p}}<m_{J/\psi}$ is always applied to eliminate the influence of the background, which corresponds to the $\psi(2S) \to X + J/\psi~(J/\psi \to p \bar{p})$ decay.
$m_{p\bar{p}}<3.077$ GeV is taken by CLEO~\cite{Alexander:2010vd}, and for BESIII, they choose $m_{p\bar{p}}<3.067$~GeV~\cite{Ablikim:2013vtm}.
Then, in the theoretical side, we can introduce the same cut to exclude the contribution of $\psi(2S) \to \eta (J/\psi \to \bar{p} p)$.
In Fig.~\ref{fig:p-p-eta}, we also give the result with the dashed-dotted red line, which includes not only the two diagrams in Fig.~\ref{fig:loop-and-background}, but also the tree diagram $\psi(2S) \to \eta (J/\psi \to \bar{p} p)$ with the $m_{p\bar{p}}<3.067$~GeV cut.
It can be seen that only the strength of the distribution becomes smaller than that in the solid black line, in which only the two diagrams in Fig.~\ref{fig:loop-and-background} are considered without any cuts, and the behaviors of these two lines are almost the same.
Thus, we can claim that the tree diagram $\psi(2S) \to \eta (J/\psi \to \bar{p} p)$ could not affect our conclusion after applying the cut to $m_{p\bar{p}}$ and it is equivalent to the statement that the Schmid theorem does not play a role on the dashed-dotted red line.
And our conclusion that there exists a visible enhancement purely caused by triangle singularity on the right shoulder of the peak structure of $N(1535)$ is still valid.

\begin{figure}[htbp]
	\centering
  \includegraphics[width=0.9\linewidth]{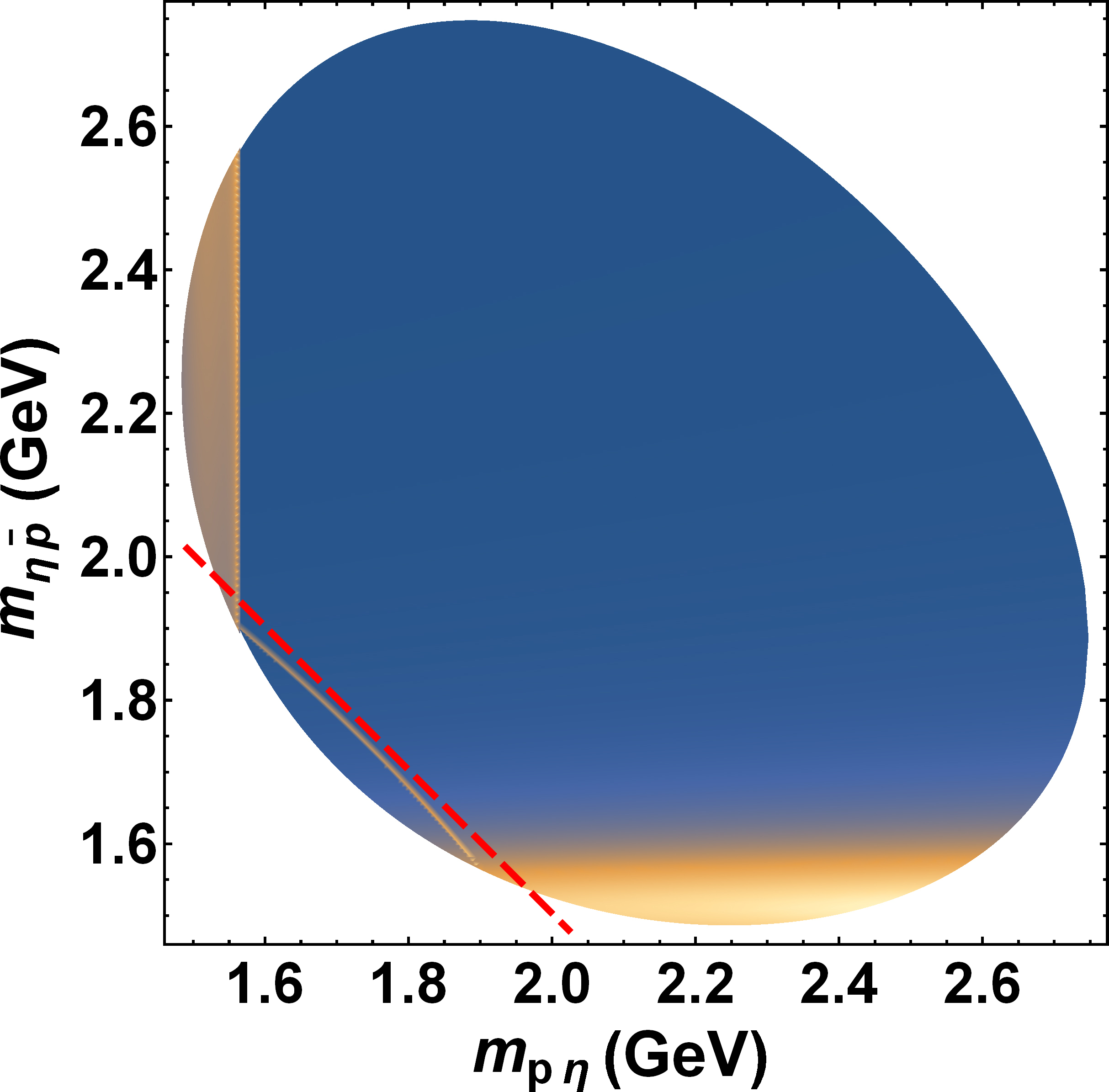}
  \caption{The Dalitz plot of the $\psi(2S) \to p \bar{p} \eta$ process after considering the contribution of Fig.~\ref{fig:loop-and-background} (a) (the vertical band), the tree diagrams $\psi(2S) \to \eta (J/\psi \to \bar{p} p)$ (the very thin band lie on the lower left corner) and $\psi(2S) \to p (\bar{N}^\ast \to \bar{p} \eta)$ (the horizontal band). For the triangle diagram, only $N(1535)$ exchange is included and $\alpha$ is set to 1. Here, the red dotted line denotes the $m_{p\bar{p}}<3.067$~GeV cut as given in Ref.~\cite{Ablikim:2013vtm}. To make all the bands visible, we increase the strength of triangle diagram by a factor $10^4$ and the $\bar{N}(1535)$ tree diagram by a factor $10^2$.}
  \label{fig:p-p-eta-dalitz}
\end{figure}

At last, we present a Dalitz plot in Fig.~\ref{fig:p-p-eta-dalitz} to explain why the three diagrams mentioned in the beginning of Sec.~\ref{sec2} can be neglected.
In Fig.~\ref{fig:p-p-eta-dalitz}, the vertical band comes entirely from the loop diagram, the horizontal band is generated by the tree diagram $\psi(2S) \to p (\bar{N}^\ast \to \bar{p} \eta)$, and the contribution of the tree diagram $\psi(2S) \to \eta (J/\psi \to \bar{p} p)$ is the very thin band just below the dashed red line, which denotes the $m_{p\bar{p}}<3.067$~GeV cut.
In general, the contribution of the loop diagram exchanging $\bar{p}$ will appear in the same region as the $\bar{N}^\ast$ tree diagram.
%
From the Daliz plot, it is clear that the contributions of the $\bar{N}^\ast$ tree diagram and the loop diagram exchanging $\bar{p}$ would not influence the triangle singularity point on the $p\eta$ invariant mass spectrum.
Besides, the contribution of the tree diagram $\psi(2S) \to \eta (J/\psi \to \bar{p} p)$ will be eliminated after applying the cut of $m_{p\bar{p}}$ as discussed above.
In a word, these three diagrams are not necessary in our calculations.

\subsection{Triangle singularity in $\psi(2S) \to p \bar{p} \pi^0$ process}

\begin{figure}[htbp]
	\centering
  \includegraphics[width=0.9\linewidth]{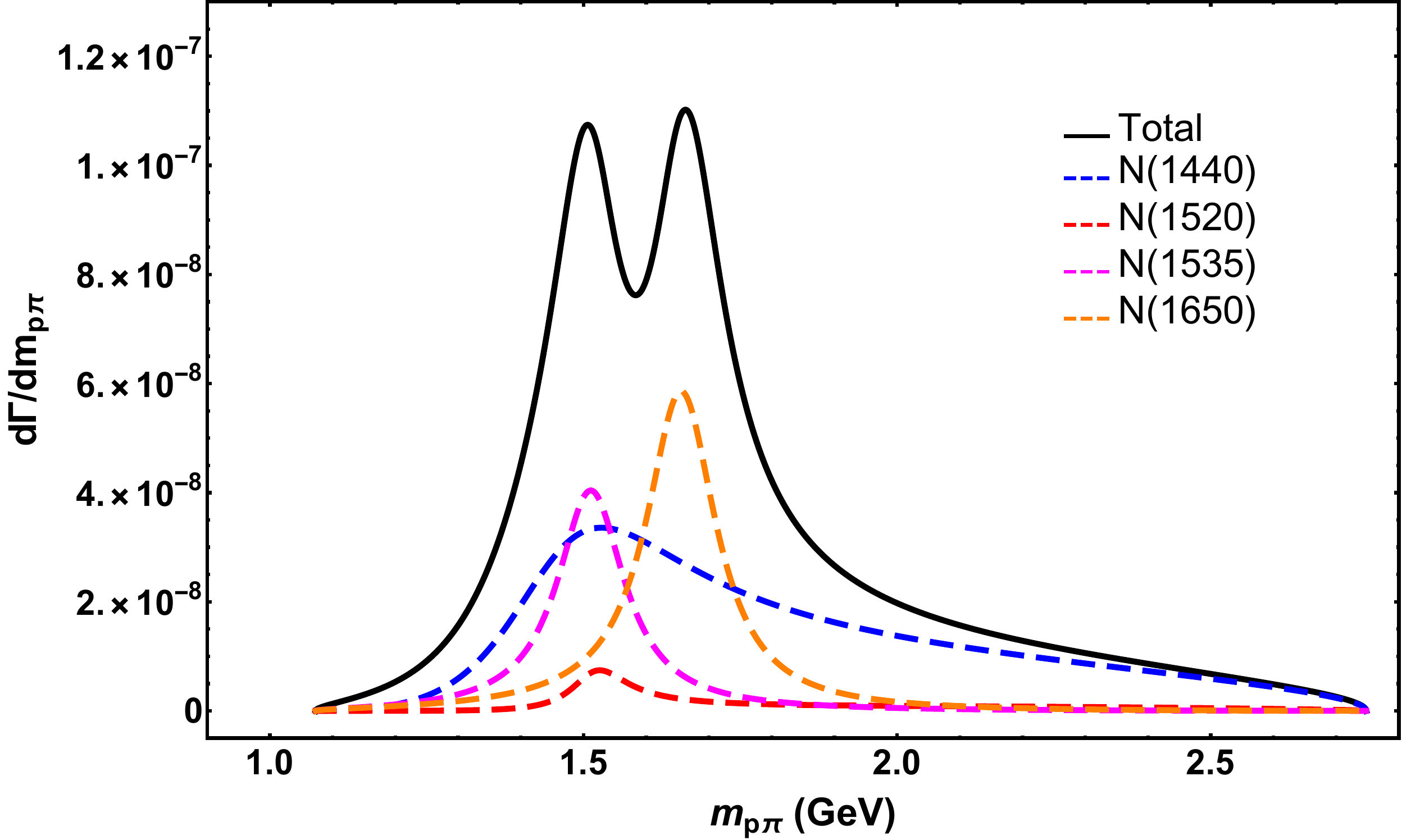}
  \caption{The tree diagram of $\psi(2S) \to p \bar{p} \pi^0$, where the contributions of $N(1440)$, $N(1520)$, $N(1535)$ and $N(1650)$ are included, and all the involved phase angles are 0.}
  \label{fig:p-p-pi-tree}
\end{figure}

Similar to our treatment on the $\psi(2S) \to p\bar{p}\eta$ process, we first prove our statement in Sec.~\ref{sec:2-Amp} that at tree level the ignorance of the phase angles between the contributions of each $N^\ast$ will not conflict with the present experimental results.
In Fig.~\ref{fig:p-p-pi-tree} we give the $m_{p\pi^0}$ distribution of the tree diagram.

After integrating $m_{p\pi^0}$, we get $\mathcal{B}_{\mathrm{Tree}}(\psi(2S) \to p \bar{p} \pi^0) = 1.68 \times 10^{-4}$, and the experimental measurement at present is $\mathcal{B}(\psi(2S) \to p \bar{p} \pi^0)= (1.65 \pm 0.03) \times 10^{-4}$ \cite{Ablikim:2012zk}. It tells that our treatment on the phase angles is acceptable.
%

\begin{figure}[htbp]
	\centering
  \includegraphics[width=0.9\linewidth]{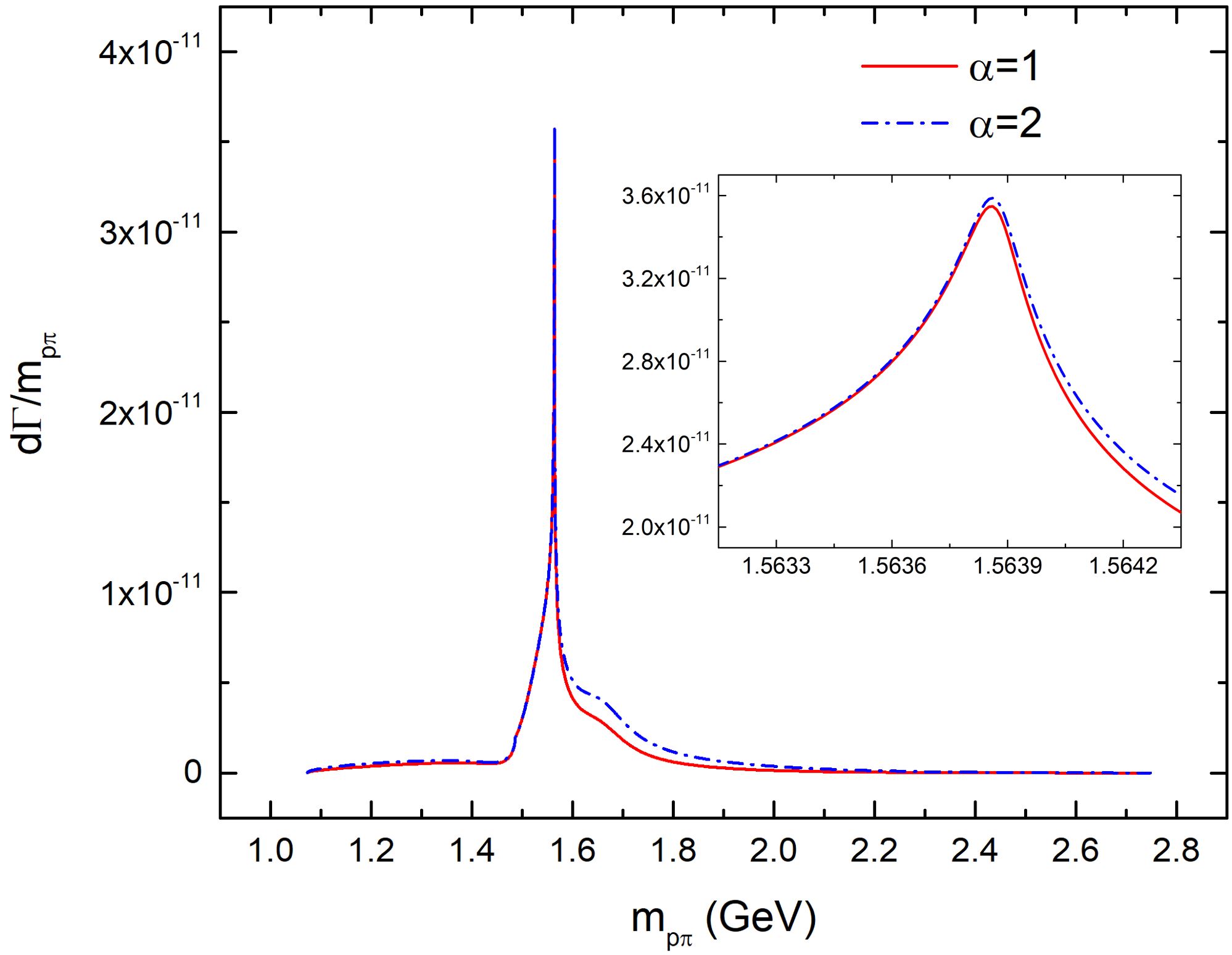}
  \caption{The $\alpha$ dependence of the triangle singularity caused by the $J/\psi \eta p$ loop.}
  \label{fig:p-p-pi-alpha}
\end{figure}

\begin{figure}[htbp]
	\centering
  \includegraphics[width=0.9\linewidth]{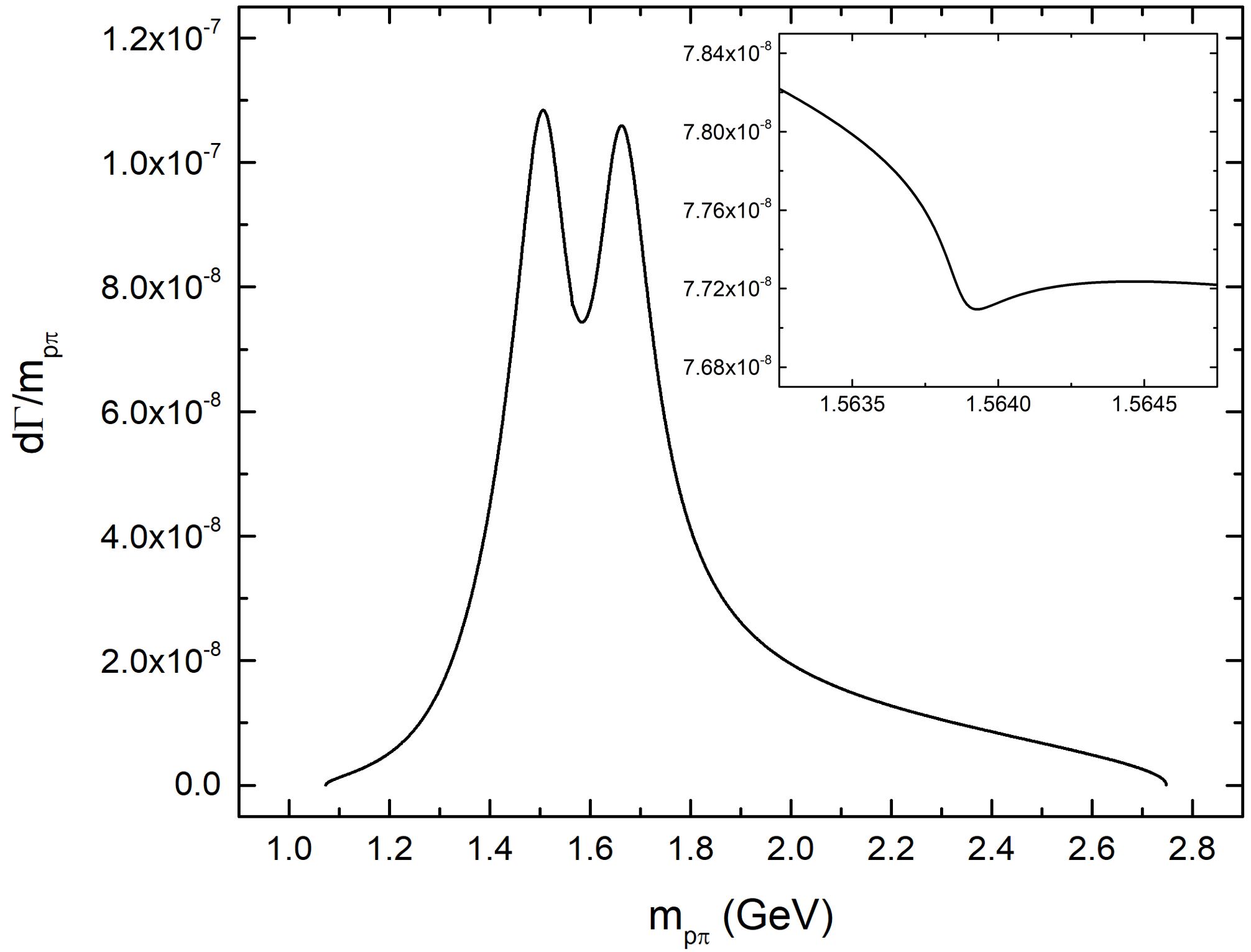}
  \caption{The $p\pi^0$ invariant mass distribution of the $\psi(2S) \to p \bar{p} \pi^0$ process after considering the interface between $\mathcal{M}^{\mathrm{Loop}}$ and $\mathcal{M}^{\mathrm{Tree}}$ (the phase angle is 0), where $\alpha$ is set to 1.}
  \label{fig:p-p-pi}
\end{figure}

Then we give the $\alpha$ dependence of loop diagram in Fig.~\ref{fig:p-p-pi-alpha}, from which we can easily see that a peak caused by the triangle singularity does appear, whose width is only about 300~keV, and it does not rely on $\alpha$ too much.
And after considering the interference between the tree and loop diagrams, the $p\pi^0$ invariant mass distribution is shown in Fig.~\ref{fig:p-p-pi}.

In Fig.~\ref{fig:p-p-pi}, we find a very small twist at the position of the triangle singularity.
After amplifying it in detail, we find a small valley structure whose width is less than 1 MeV.
The detraction caused by this valley is only about $1\%$ from the calculation.
If there exists 4 billion $\psi(2S)$ events, the event corresponding to this effect is only 10, which implies that it would be impossible to observe experimentally.

It is understandable that the triangle singularity effect in the $p\bar{p}\pi^0$ case is much smaller than that in the $p\bar{p}\eta$ case.
As shown in Figs.~\ref{fig:p-p-eta-alpha} and \ref{fig:p-p-pi-alpha}, the strength of the sharp peak from the pure triangle singularity in the $p\eta$ case is much higher than that in the $p\pi$ case and the difference between them is about 20 times.
The reason is that the branching ratio of $N(1535)$ to the $p\eta$ final state is larger than that to the $p\pi^0$ final state with a factor 6 including the isospin factor.
Furthermore, the interference between the background including $N(1650)$ and $N(1535)$ for $p\pi$ channel would weaken the contribution of $N(1535)$ as shown in Fig.~\ref{fig:cross-section-fit}.
As a result, the pure triangle singularity effect is suppressed in the $p\bar{p}\pi^0$ process.

\subsection{Further discussions on how to strengthen the triangle singularity effect}

We have mentioned that the smaller the widths of the intermediate exchange particles are, the sharper the peak of triangular singularity is.
However, through a detailed calculation, we find that if the widths of the intermediate particles in the loop are too small, some other problems may arise.
In the cases of this work, where the internal particles are $J/\psi$, $\eta$ and $p$, whose widths are 92.9 keV, 1.31 keV and 0, respectively~\cite{Zyla:2020zbs}.
The result given above tells that the width of the pure triangle singularity is only about 1~MeV and it is enlarged to 5~MeV after including the interference with the tree diagram,.
From the parameters of BESIII~\cite{Ablikim:2019hff}, we can know that the resolution of BESIII experiment is about 4.3~MeV. Hence it is almost impossible to observe this structure from the BESIII detection currently, unless BESIII or other experiments, such as STCF, can improve their resolutions to 2-3~MeV in the future~\cite{Discussion:Lyu}.

\begin{figure}[htbp]
	\centering
  \includegraphics[width=0.9\linewidth]{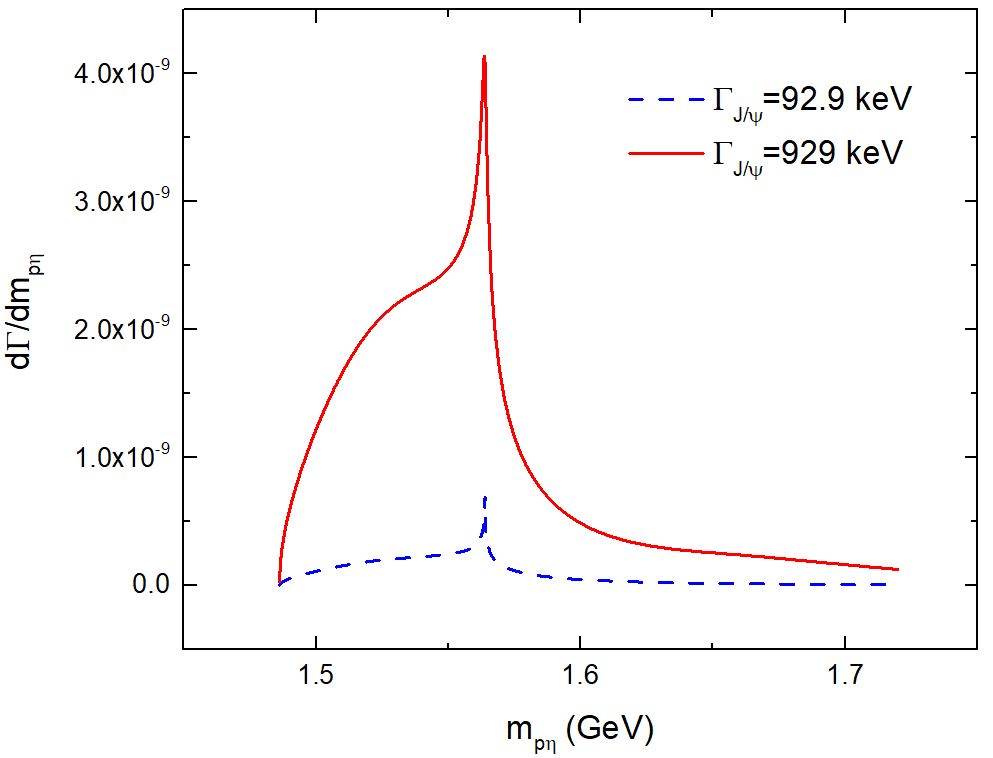}
  \caption{The peaks caused by the triangle singularity with different assumptions of the width of $J/\psi$.}
  \label{fig:p-p-eta-psi}
\end{figure}

In addition to the experimental observation problems, another problem caused by the small widths is that the intensity of the peak will be weakened.
We still consider the $J/\psi \eta p$ loop case as an example. As claimed previously, when the triangle singularity happens, all the involved particles are on mass-shell, so the coupling constants of interaction vertices can be extracted from the corresponding decay processes, for example, $g_{J/\psi p \bar{p}}$ is computed from the partial decay width of $J/\psi$ to $p \bar{p}$.
Then we assume the branching ratio of $J/\psi$ to $p \bar{p}$ is fixed and increase the total width of $J/\psi$ from 92.9~keV to 929~keV. The peaks caused by the pure triangle singularity with these two different assumptions of the width of $J/\psi$ is presented in Fig.~\ref{fig:p-p-eta-psi}.

From Fig.~\ref{fig:p-p-eta-psi} we can see that when we enlarge the width of $J/\psi$, the peak of the triangle singularity not only becomes wider, but also becomes higher.
The reason for broadening this peak is discussed in Ref.~\cite{Du:2019idk}.
Here we try to figure out why the strength of triangle singularity is also related with the widths of internal particles.
Let's consider the simplest loop integral which only involves the participated coupling constant, form factors and the propagators of internal particles as an example,
\begin{eqnarray}
  \mathcal{I}&\equiv& g\int\frac{d^4 q}{(2\pi)^4} \frac{\mathcal{F}(p_3+p_4-q, m_\eta, \Lambda_\eta)} {(p_3+p_4-q)^2-m_\eta^2+im_\eta\Gamma_\eta}\nonumber\\
  &&\times \frac{\mathcal{F}(p_2+q,m_{J/\psi},\Lambda_{J/\psi})} {(p_2+q)^2-m_{J/\psi}^2+im_{J/\psi}\Gamma_{J/\psi}}\nonumber\\
  &&\times\frac{\mathcal{F} (q,m_p,\Lambda_p)} {q^2-m_p^2+i m_p \Gamma_p},\label{eq:loop-i-4}
\end{eqnarray}
with $g$ being the constant that scales the coupling constant $g_{J/\psi p p}$, since $\mathcal{B}(J/\psi \to p \bar{p})$ is fixed in our assumption.
Actually, the triangle singularity is divergent if all the widths of the exchanged particles are zero.
There is a very sharp structure when $\cos \theta = -1$, where $\theta$ is the relative angle between the momentums of internal $J/\psi$ and outgoing $\bar{p}$.
Then we consider the relationship of $\mathcal{I}$ and $\cos\theta$ and choose the definition as
\begin{eqnarray}
  I:~\mathcal{I}=\int d \cos\theta I(\cos\theta).\label{eq:loop-i-3}
\end{eqnarray}

\begin{figure}[htbp]
	\centering
  \includegraphics[width=0.9\linewidth]{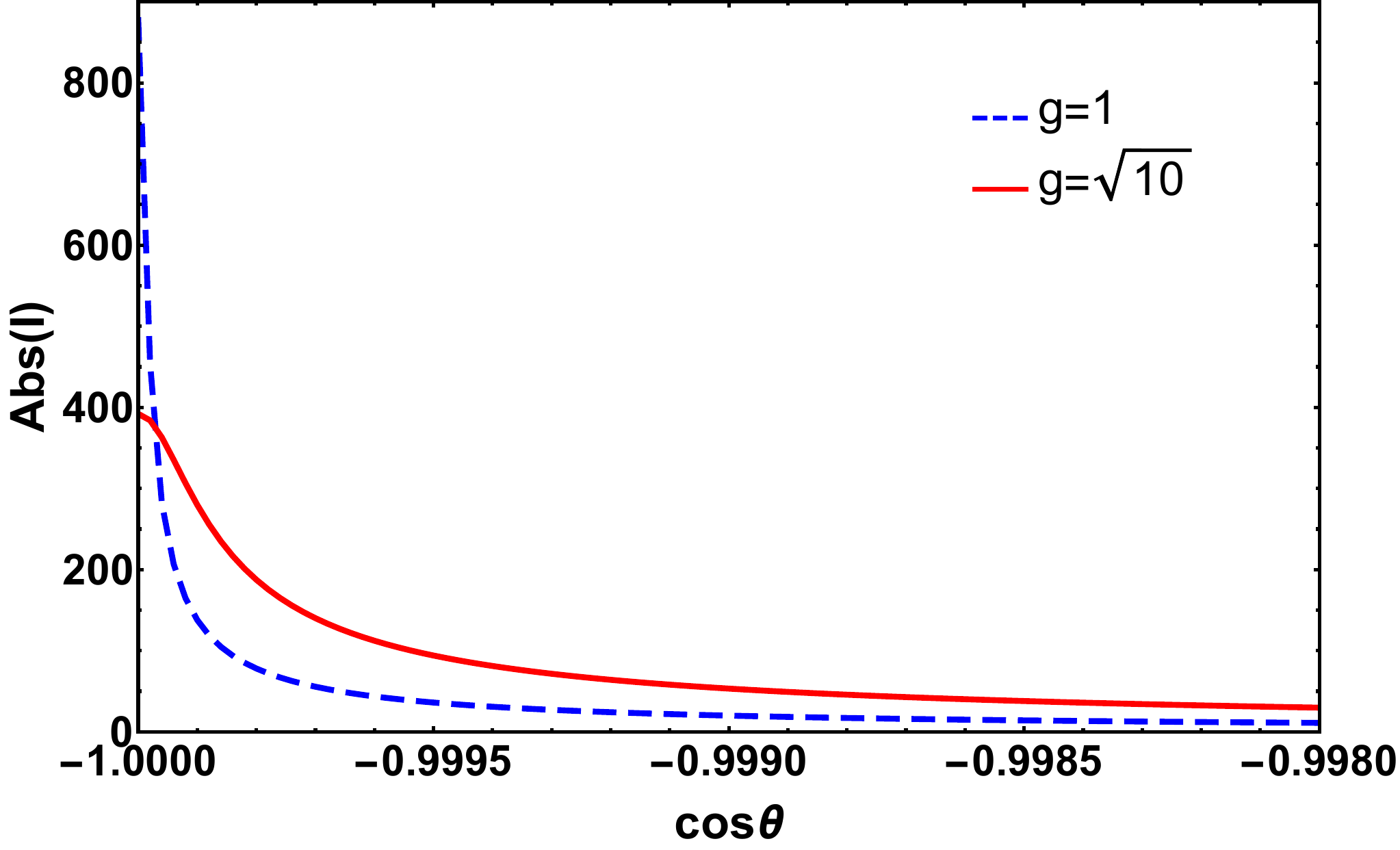}
  \caption{The dependence behavior of $I$ on $cos \theta$ for different values of $g$.}
  \label{fig:p-p-eta-I}
\end{figure}

The dependence behaviors of $I$ on $\cos \theta$ for different values of $\Gamma_{J/\psi}$ are given in Fig.~\ref{fig:p-p-eta-I}.
Here we set $(p_3+p_4)^2=1.56387^2~\mathrm{GeV}^2$ to make it being just the triangle singularity point.
We draw the result around $\cos \theta = -1$, since we only interested in the region around the triangle singularity.
When the width of $J/\psi$ changes from 92.9~keV to 929~keV, the coupling constant $g_{J/\psi p p}$ will certainly become larger, and in this case the scale factor $g$ goes from 1 to $\sqrt{10}$.
First of all, we can see that there is a 2 times difference in the value of $I$ for these two cases at $\cos \theta = -1$.
Since $|\vec{q}|$ is integrated, the off-shell effect of $J/\psi$ will influence the value of $I$ even at the point of $\cos \theta = -1$.
It leads to the fact that although the decrease factor 2 is close, it is still less than the factor $\sqrt{10}$, which is the precise value for the on-shell case, i.e., the ratios of $(g/\Gamma_{J/\psi})$ in two cases of $J/\psi$'s widths. 
Once $\cos \theta$ is greater than $-1$, particle $J/\psi$ is totally off-shell, which makes $(p_2+q)^2-m_{J/\psi}^2$ become dominant in the denominate of $J/\psi$'s propagator.
And here the ratio between red and blue line will be close to $\sqrt{10}$, which is from purely the change of $g$.
Obviously, the value of $\mathcal{I}$ gets larger after increasing the width of $J/\psi$.
Finally, after we integrate $\cos\theta$, we can get $\mathcal{I}=0.154$ and $0.360$ for $\Gamma_{J/\psi}=92.9$~keV and $929$~keV, respectively.
In other words, when the width of $J/\psi$ increases, the loop function is enhanced, which leads to the enhancement of the contribution of pure singularity.
On the other hand, although the absolute value at the peak of the triangle singularity increases, the width of this peak is also increasing, and it is even worse that the difference between the peak value and the background value becomes smaller as shown in Fig.~\ref{fig:p-p-eta-psi}.
That is, the significance of the triangle singularity decreases just as the statement in the introduction.

In short, the widths of the intermediate particles should not be too small to observe this pure triangle singularity effect easily in experiment.
Actually, there exists a balance for the widths of the internal particles.
When the widths are too large, the peak becomes very wide and insignificant, and it might be mixed with the threshold and resonance states, so it is difficult to distinguish the pure contributions.
On the contrary, when the widths are very small, the structure is very narrow and the intensity is very weak, which makes it difficult to observe experimentally.
Therefore, it is necessary to find processes, whose intermediate particles have appropriate widths, to help experiments to search for the pure triangle singularity effect.
And we will focus on this idea in our further research.

\section{Summary}

The triangle singularity proposed by L. D. Landau ~\cite{Landau:1959fi} might be very important in explaining many experimental results ~\cite{BESIII:2012aa,Ablikim:2013mio,Liu:2013dau,Xiao:2013iha,Ablikim:2013wzq,Ablikim:2013xfr,Ablikim:2013emm,Ablikim:2017oaf,Aaij:2015tga,Aaij:2019vzc}.
However, up to now, an effect purely caused by the triangle singularity is still absent, which means that the triangle singularity itself has not been confirmed by experimental observations.
There are three main reasons for this. Firstly, the triangle singularity is always mixed with the threshold~\cite{Liu:2015taa}, secondly, the widths of the intermediate particles are large~\cite{Achasov:2015uua,Du:2019idk}, and finally, it is due to the lack of information of each vertex.
Thus, aiming to observe a pure triangle singularity effect, the position of the triangle singularity should be away from the threshold and the widths of the particles that composed the loop should be as small as possible, which leads the peak to be sharp enough so that it can be distinguished, furthermore, it would be much better if all the vertices can be constrainted precisely.

Guided by these ideas, in this work we propose to observe a pure triangle singularity effect in the $\psi(2S) \to p \bar{p} \eta / p\bar{p} \pi^0$ processes, where the triangle loop is composed by $J/\psi$, $\eta$ and $p$.
After applying Coleman-Norton theorem, we get the position of triangle singularity in $p\eta(\pi^0)$ invariant mass spectrum is 1.56387~GeV, which is about 80~MeV away from the $p \eta$ threshold.
In addition, since all the widths of these particles are very small, the peak caused by the triangle singularity must be very sharp.
Moreover, all the involved vertices can be extracted from experimental data precisely.
Therefore, the triangle singularity in the $\psi(2S) \to p \bar{p} \eta / p\bar{p} \pi^0$ processes will not mix up with the $p \eta$ threshold and it can be distinguished easily from $N^\ast$ resonances since their widths are around 100~MeV~\cite{Zyla:2020zbs}, meanwhile, it can be predicted precisely in theory.

From our numerical results, there exists a $10\%$ enhancement around 1.56387~GeV compared with the background in the invariant mass spectrum of $p\eta$ for the $\psi(2S) \to p \bar{p} \eta$ process, and the width of this peak is about 5~MeV.
There will be about 120 events for this enhancement when there are 4 billion $\psi(2S)$ events.
The only assumption we made is that the interference phase angle between the tree and loop diagrams is zero, and the significance of the peak would change along with this phase angle.
While for the $p \bar{p} \pi^0$ case, the significance of the triangle singularity is very small and could be negligible experimentally.
At last, after the discussion with experimentalists~\cite{Discussion:Lyu} we find that to distinguish such a narrow triangle singularity peak needs a high resolution around 2-3~MeV. While unfortunately, the BESIII detection can not satisfy this requirement currently.
We expect the future facilities such as STCF can make significant improvements on the resolution.

Since triangle singularity is a pure kinematical effect, it is not dependent on the dynamics of particles, i.e., triangle singularity is a model independent effect.
It tells that our prediction is very precise and can be compared with the experimental data directly.
Therefore, we suggest experiments such as BESIII and STCF to make a precise analysis on the $p\eta$ invariant mass spectrum of the $\psi(2S) \to p \bar{p} \eta$ process.
If this small narrow peak is observed, it will be the first time that a pure triangle singularity effect is observed, which will not only help us understand the triangle singularity itself, but also be a real discovery of hadron loop mechanism.

\section*{Acknowledgments}

The authors want to thank Feng-kun Guo, Satoshi Nakamura, Eulogio Oset, Qiang Zhao, and Bing-Song Zou for useful discussions.
We are grateful to experimentalists, Bei-jiang Liu, Xiao-Rui Lyu and Zi-Yi Wang, who help us to understand the limitation of detection experimentally.
This work is supported in part by the Fundamental Research Funds for the Central Universities.
\vfil


\begin{thebibliography}{333}


\bibitem{Landau:1959fi}
L.~D.~Landau,
On analytic properties of vertex parts in quantum field theory,
Nucl. Phys. \textbf{13}, no.1, 181-192 (1960).

\bibitem{BESIII:2012aa}
M.~Ablikim \textit{et al.} [BESIII],
First observation of $\eta(1405)$ decays into $f_{0}(980)\pi^0$,
Phys. Rev. Lett. \textbf{108}, 182001 (2012).

\bibitem{Wu:2011yx}
J.~J.~Wu, X.~H.~Liu, Q.~Zhao and B.~S.~Zou,
The Puzzle of anomalously large isospin violations in $\eta(1405/1475)\to 3\pi$,
Phys. Rev. Lett. \textbf{108}, 081803 (2012).

\bibitem{Aceti:2012dj}
F.~Aceti, W.~H.~Liang, E.~Oset, J.~J.~Wu and B.~S.~Zou,
Isospin breaking and $f_0(980)$-$a_0(980)$ mixing in the $\eta(1405) \to \pi^{0} f_0(980)$ reaction,
Phys. Rev. D \textbf{86}, 114007 (2012).

\bibitem{Wu:2012pg}
X.~G.~Wu, J.~J.~Wu, Q.~Zhao and B.~S.~Zou,
Understanding the property of $\eta(1405/1475)$ in the $J/\psi$ radiative decay,
Phys. Rev. D \textbf{87}, no.1, 014023 (2013).


\bibitem{Achasov:2015uua}
N.~N.~Achasov, A.~A.~Kozhevnikov and G.~N.~Shestakov,
``Isospin breaking decay $\eta(1405) \to f_0(980)\pi^0 \to 3\pi$,''
Phys. Rev. D \textbf{92} (2015) no.3, 036003

\bibitem{Du:2019idk}
M.~C.~Du and Q.~Zhao,
``Internal particle width effects on the triangle singularity mechanism in the study of the $\eta(1405)$ and $\eta(1475)$ puzzle,''
Phys. Rev. D \textbf{100} (2019) no.3, 036005


\bibitem{Ketzer:2015tqa}
M.~Mikhasenko, B.~Ketzer and A.~Sarantsev,
Nature of the $a_1(1420)$,
Phys. Rev. D \textbf{91}, no.9, 094015 (2015).


\bibitem{Ablikim:2013mio}
  M.~Ablikim {\it et al.} [BESIII Collaboration],
  Observation of a Charged Charmoniumlike Structure in $e^+e^- \to J/\psi \pi^+ \pi^-$ at $\sqrt{s}$ =4.26 GeV,
  Phys.\ Rev.\ Lett.\  {\bf 110}, 252001 (2013)

\bibitem{Liu:2013dau}
  Z.~Q.~Liu {\it et al.} [Belle Collaboration],
  Study of $e^+e^- \to \pi^+ \pi^- J/\psi$ and Observation of a Charged Charmoniumlike State at Belle,
  Phys.\ Rev.\ Lett.\  {\bf 110}, 252002 (2013)

\bibitem{Xiao:2013iha}
  T.~Xiao, S.~Dobbs, A.~Tomaradze and K.~K.~Seth,
  Observation of the Charged Hadron $Z_c^{\pm}(3900)$ and Evidence for the Neutral $Z_c^0(3900)$ in $e^+e^-\to \pi\pi J/\psi$ at $\sqrt{s}=4170$ MeV,
  Phys.\ Lett.\ B {\bf 727}, 366 (2013)

\bibitem{Ablikim:2013wzq}
  M.~Ablikim {\it et al.} [BESIII Collaboration],
  Observation of a Charged Charmoniumlike Structure $Z_c$(4020) and Search for the $Z_c$(3900) in $e^+e^- \to \pi^+ \pi^- h_c$,
  Phys.\ Rev.\ Lett.\  {\bf 111}, no. 24, 242001 (2013)

\bibitem{Ablikim:2013xfr}
  M.~Ablikim {\it et al.} [BESIII Collaboration],
  Observation of a charged $(D\bar{D}^{*})^\pm$ mass peak in $e^{+}e^{-} \to \pi D\bar{D}^{*}$ at $\sqrt{s} =$ 4.26 GeV,
  Phys.\ Rev.\ Lett.\  {\bf 112}, no. 2, 022001 (2014)

\bibitem{Ablikim:2013emm}
  M.~Ablikim {\it et al.} [BESIII Collaboration],
  Observation of a charged charmoniumlike structure in $e^+e^- \to (D^{*} \bar{D}^{*})^{\pm} \pi^\mp$ at $\sqrt{s}=4.26$GeV,
  Phys.\ Rev.\ Lett.\  {\bf 112}, no. 13, 132001 (2014)

\bibitem{Ablikim:2017oaf}
  M.~Ablikim {\it et al.} [BESIII Collaboration],
  Measurement of $e^{+}e^{-}\rightarrow \pi^{+}\pi^{-}\psi(3686)$ from 4.008 to 4.600~GeV and observation of a charged structure in the $\pi^{\pm}\psi(3686)$ mass spectrum,
  Phys.\ Rev.\ D {\bf 96}, no. 3, 032004 (2017)

\bibitem{Aaij:2015tga}
R.~Aaij \textit{et al.} [LHCb],
Observation of $J/\psi p$ Resonances Consistent with Pentaquark States in $\Lambda_b^0 \to J/\psi K^- p$ Decays,
Phys. Rev. Lett. \textbf{115}, 072001 (2015).

\bibitem{Aaij:2019vzc}
R.~Aaij \textit{et al.} [LHCb],
Observation of a narrow pentaquark state, $P_c(4312)^+$, and of two-peak structure of the $P_c(4450)^+$,'
Phys. Rev. Lett. \textbf{122}, no.22, 222001 (2019).

\bibitem{Wang:2013cya}
Q.~Wang, C.~Hanhart and Q.~Zhao,
Decoding the riddle of $Y(4260)$ and $Z_c(3900)$,
Phys. Rev. Lett. \textbf{111}, no.13, 132003 (2013).

\bibitem{Wang:2013hga}
Q.~Wang, C.~Hanhart and Q.~Zhao,
Systematic study of the singularity mechanism in heavy quarkonium decays,
Phys. Lett. B \textbf{725}, no.1-3, 106-110 (2013).


\bibitem{Liu:2015taa}
X.~H.~Liu, M.~Oka and Q.~Zhao,
Searching for observable effects induced by anomalous triangle singularities,
Phys. Lett. B \textbf{753}, 297-302 (2016).

\bibitem{Liu:2015fea}
X.~H.~Liu, Q.~Wang and Q.~Zhao,
Understanding the newly observed heavy pentaquark candidates,
Phys. Lett. B \textbf{757}, 231-236 (2016).

\bibitem{Guo:2015umn}
F.~K.~Guo, U.~G.~Mei\ss{}ner, W.~Wang and Z.~Yang,
How to reveal the exotic nature of the P$_c$(4450),
Phys. Rev. D \textbf{92}, no.7, 071502 (2015).

\bibitem{Szczepaniak:2015eza}
A.~P.~Szczepaniak,
Triangle Singularities and XYZ Quarkonium Peaks,
Phys. Lett. B \textbf{747}, 410-416 (2015).

\bibitem{Guo:2016bkl}
F.~K.~Guo, U.~G.~Mei\ss{}ner, J.~Nieves and Z.~Yang,
Remarks on the $P_c$ structures and triangle singularities,
Eur. Phys. J. A \textbf{52}, no.10, 318 (2016).

\bibitem{Bayar:2016ftu}
M.~Bayar, F.~Aceti, F.~K.~Guo and E.~Oset,
A Discussion on Triangle Singularities in the $\Lambda_b \to J/\psi K^{-} p$ Reaction,
Phys. Rev. D \textbf{94}, no.7, 074039 (2016).

\bibitem{Wang:2016dtb}
E.~Wang, J.~J.~Xie, W.~H.~Liang, F.~K.~Guo and E.~Oset,
Role of a triangle singularity in the $\gamma p\rightarrow K^+ \Lambda(1405)$ reaction,
Phys. Rev. C \textbf{95}, no.1, 015205 (2017).

\bibitem{Pilloni:2016obd}
A.~Pilloni \textit{et al.} [JPAC],
Amplitude analysis and the nature of the Z$_c$(3900),
Phys. Lett. B \textbf{772}, 200-209 (2017).

\bibitem{Xie:2016lvs}
J.~J.~Xie, L.~S.~Geng and E.~Oset,
$f_2$(1810) as a triangle singularity,
Phys. Rev. D \textbf{95}, no.3, 034004 (2017).

\bibitem{Szczepaniak:2015hya}
A.~P.~Szczepaniak,
Dalitz plot distributions in presence of triangle singularities,
Phys. Lett. B \textbf{757}, 61-64 (2016).

\bibitem{Roca:2017bvy}
L.~Roca and E.~Oset,
Role of a triangle singularity in the $\pi \Delta$ decay of the $N(1700)(3/2^-)$,
Phys. Rev. C \textbf{95}, no.6, 065211 (2017).

\bibitem{Debastiani:2017dlz}
V.~R.~Debastiani, S.~Sakai and E.~Oset,
Role of a triangle singularity in the $\pi N(1535)$ contribution to $\gamma p \to p \pi^0 \eta$,
Phys. Rev. C \textbf{96}, no.2, 025201 (2017).

\bibitem{Samart:2017scf}
D.~Samart, W.~h.~Liang and E.~Oset,
Triangle mechanisms in the build up and decay of the $N^*(1875)$,
Phys. Rev. C \textbf{96}, no.3, 035202 (2017).

\bibitem{Sakai:2017hpg}
S.~Sakai, E.~Oset and A.~Ramos,
Triangle singularities in $B^-\rightarrow K^-\pi^-D_{s0}^+$ and $B^-\rightarrow K^-\pi^-D_{s1}^+$,
Eur. Phys. J. A \textbf{54}, no.1, 10 (2018).

\bibitem{Pavao:2017kcr}
R.~Pavao, S.~Sakai and E.~Oset,
Triangle singularities in $B^-\rightarrow D^{*0}\pi ^-\pi ^0\eta $ and $B^-\rightarrow D^{*0}\pi ^-\pi ^+\pi ^-$,
Eur. Phys. J. C \textbf{77}, no.9, 599 (2017).

\bibitem{Xie:2017mbe}
J.~J.~Xie and F.~K.~Guo,
Triangular singularity and a possible $\phi p$ resonance in the $\Lambda^+_c \to \pi^0 \phi p$ decay,
Phys. Lett. B \textbf{774}, 108-113 (2017).

\bibitem{Bayar:2017svj}
M.~Bayar, R.~Pavao, S.~Sakai and E.~Oset,
Role of the triangle singularity in $\Lambda(1405)$ production in the $\pi^-p\rightarrow K^0\pi\Sigma$ and $pp\rightarrow pK^+\pi\Sigma$ processes,
Phys. Rev. C \textbf{97}, no.3, 035203 (2018).

\bibitem{Liang:2017ijf}
W.~H.~Liang, S.~Sakai, J.~J.~Xie and E.~Oset,
Triangle singularity enhancing isospin violation in $\bar B_s^0 \to J/\psi \pi^0 f_0(980)$,
Chin. Phys. C \textbf{42}, no.4, 044101 (2018).

\bibitem{Oset:2018zgc}
E.~Oset and L.~Roca,
Triangle singularity in $\tau \to f_1(1285)\pi\nu_\tau$ decay,
Phys. Lett. B \textbf{782}, 332-338 (2018).

\bibitem{Dai:2018hqb}
L.~R.~Dai, R.~Pavao, S.~Sakai and E.~Oset,
Anomalous enhancement of the isospin-violating $\Lambda(1405)$ production by a triangle singularity in $\Lambda_c\rightarrow\pi^+\pi^0\pi^0\Sigma^0$,
Phys. Rev. D \textbf{97}, no.11, 116004 (2018).

\bibitem{Dai:2018rra}
L.~R.~Dai, Q.~X.~Yu and E.~Oset,
Triangle singularity in $\tau^- \to \nu_\tau \pi^- f_0(980)$ ($a_0(980)$) decays,
Phys. Rev. D \textbf{99}, no.1, 016021 (2019).

\bibitem{Liu:2020orv}
X.~H.~Liu, M.~J.~Yan, H.~W.~Ke, G.~Li and J.~J.~Xie,
Triangle singularity as the origin of $X_0(2900)$ and $X_1(2900)$ observed in $B^+\to D^+ D^- K^+$,

\bibitem{Guo:2019qcn}
F.~K.~Guo,
Novel Method for Precisely Measuring the $X(3872)$ Mass,
Phys. Rev. Lett. \textbf{122}, no.20, 202002 (2019).

\bibitem{Liang:2019jtr}
W.~H.~Liang, H.~X.~Chen, E.~Oset and E.~Wang,
Triangle singularity in the $J/\psi \rightarrow K^+ K^- f_0(980)(a_0(980))$ decays,
Eur. Phys. J. C \textbf{79}, no.5, 411 (2019).

\bibitem{Nakamura:2019emd}
S.~X.~Nakamura,
Triangle singularities in $\bar{B}^0\to \chi_{c1}K^-\pi^+$ relevant to $Z_1(4050)$ and $Z_2(4250)$,
Phys. Rev. D \textbf{100}, no.1, 011504 (2019).

\bibitem{Liu:2019dqc}
X.~H.~Liu, G.~Li, J.~J.~Xie and Q.~Zhao,
Visible narrow cusp structure in $\Lambda_c^+\to p K^- \pi^+$ enhanced by triangle singularity,
Phys. Rev. D \textbf{100}, no.5, 054006 (2019).

\bibitem{Jing:2019cbw}
H.~J.~Jing, S.~Sakai, F.~K.~Guo and B.~S.~Zou,
Triangle singularities in ${J/\psi\rightarrow\eta\pi^0\phi}$ and ${\pi^0\pi^0\phi}$,
Phys. Rev. D \textbf{100}, no.11, 114010 (2019).

\bibitem{Braaten:2019gfj}
E.~Braaten, L.~P.~He and K.~Ingles,
Triangle Singularity in the Production of X(3872) and a Photon in $e^+e^-$ Annihilation,
Phys. Rev. D \textbf{100}, no.3, 031501 (2019).

\bibitem{Sakai:2020ucu}
S.~Sakai, E.~Oset and F.~K.~Guo,
Triangle singularity in the $B^-\to K^-\pi^0X(3872)$ reaction and sensitivity to the $X(3872)$ mass,
Phys. Rev. D \textbf{101}, no.5, 054030 (2020).

\bibitem{Sakai:2020fjh}
S.~Sakai,
Role of the triangle mechanism in the $\Lambda_b\rightarrow \Lambda_c\pi^-f_0(980)$ reaction,
Phys. Rev. D \textbf{101}, no.7, 074041 (2020).

\bibitem{Molina:2020kyu}
R.~Molina and E.~Oset,
Triangle singularity in $B^-\rightarrow K^-X(3872);X\rightarrow \pi ^0\pi ^+\pi ^-$ and the X(3872) mass,
Eur. Phys. J. C \textbf{80}, no.5, 451 (2020).

\bibitem{Braaten:2020iye}
E.~Braaten, L.~P.~He, K.~Ingles and J.~Jiang,
Charm-meson triangle singularity in ${e^+e^-}$ annihilation into ${ D^{*0} \bar{D}^0 + \gamma }$,
Phys. Rev. D \textbf{101}, no.9, 096020 (2020).

\bibitem{Alexeev:2020lvq}
M.~G.~Alexeev \textit{et al.} [COMPASS],
A Triangle Singularity as the Origin of the $a_1(1420)$,

\bibitem{Ortega:2020ayw}
P.~G.~Ortega and E.~Ruiz Arriola,
On the precise measurement of the $X(3872)$ mass and its counting rate,

\bibitem{Shen:2020gpw}
C.~W.~Shen, H.~J.~Jing, F.~K.~Guo and J.~J.~Wu,
Exploring possible triangle singularities in the $\Xi^-_{b} \to K^- J/\psi \Lambda$ decay,
Symmetry \textbf{12}, no.10, 1611 (2020).

\bibitem{Achasov:2019wvw}
N.~N.~Achasov and G.~N.~Shestakov,
Decay $X(3872)\to\pi^0\pi^+\pi^-$ and $S$-wave $D^0\bar D^0 \to\pi^+\pi^-$ scattering length,
Phys. Rev. D \textbf{99}, no.11, 116023 (2019).

\bibitem{Guo:2019twa}
F.~K.~Guo, X.~H.~Liu and S.~Sakai,
Threshold cusps and triangle singularities in hadronic reactions,
Prog. Part. Nucl. Phys. \textbf{112}, 103757 (2020).

\bibitem{Aaij:2020hon}
R.~Aaij \textit{et al.} [LHCb],
A model-independent study of resonant structure in $B^+\to D^+D^-K^+$ decays,

\bibitem{Aaij:2020ypa}
R.~Aaij \textit{et al.} [LHCb],
Amplitude analysis of the $B^+\to D^+D^-K^+$ decay,

\bibitem{Tsushima:1996xc}
K.~Tsushima, A.~Sibirtsev and A.~W.~Thomas,
Resonance model study of strangeness production in p p collisions,
Phys. Lett. B \textbf{390}, 29-35 (1997).

\bibitem{Tsushima:1998jz}
K.~Tsushima, A.~Sibirtsev, A.~W.~Thomas and G.~Q.~Li,
Resonance model study of kaon production in baryon baryon reactions for heavy ion collisions,
Phys. Rev. C \textbf{59}, 369-387 (1999)
[erratum: Phys. Rev. C \textbf{61}, 029903 (2000)]

\bibitem{Zou:2002yy}
B.~S.~Zou and F.~Hussain,
Covariant L-S scheme for the effective N*NM couplings,
Phys. Rev. C \textbf{67}, 015204 (2003).

\bibitem{Ouyang:2009kv}
Z.~Ouyang, J.~J.~Xie, B.~S.~Zou and H.~S.~Xu,
Theoretical study on $p p \to p n \pi^+$ reaction at medium energies,
Int. J. Mod. Phys. E \textbf{18}, 281-292 (2009).

\bibitem{Wu:2009md}
J.~J.~Wu, Z.~Ouyang and B.~S.~Zou,
Proposal for Studying N* Resonances with $\bar{p} p \to \bar{p} n \pi^+$ Reaction,
Phys. Rev. C \textbf{80}, 045211 (2009).

\bibitem{Cao:2010km}
X.~Cao, B.~S.~Zou and H.~S.~Xu,
Phenomenological analysis of the double pion production in nucleon-nucleon collisions up to 2.2 GeV,
Phys. Rev. C \textbf{81}, 065201 (2010).

\bibitem{Cao:2010ji}
X.~Cao, B.~S.~Zou and H.~S.~Xu,
Phenomenological study on the $\bar p N\to \bar NN\pi\pi$ reactions,
Nucl. Phys. A \textbf{861}, 23-36 (2011).

\bibitem{Xu:2015qqa}
H.~Xu, J.~J.~Xie and X.~Liu,
Implication of the observed $e^{+}e^{-}\rightarrow p{\bar{p}}\pi ^0$ for studying the $p{\bar{p}}\rightarrow \psi (3770)\pi ^0$ process,
Eur. Phys. J. C \textbf{76}, no.4, 192 (2016).

\bibitem{Zyla:2020zbs}
P.~A.~Zyla \textit{et al.} [Particle Data Group],
Review of Particle Physics,
PTEP \textbf{2020}, no.8, 083C01 (2020).

\bibitem{Prakhov:2005qb}
S.~Prakhov, B.~M.~K.~Nefkens, C.~E.~Allgower, R.~A.~Arndt, V.~Bekrenev, W.~J.~Briscoe, M.~Clajus, J.~R.~Comfort, K.~Craig and D.~Grosnick, \textit{et al.}
Measurement of $\pi^- p \to \eta n$ from threshold to p ($\pi^-$) = 747 MeV,
Phys. Rev. C \textbf{72}, 015203 (2005).

\bibitem{Deinet:1969cd}
W.~Deinet, H.~Mueller, D.~Schmitt, H.~M.~Staudenmaier, S.~Buniatov and E.~Zavattini,
Differential and total cross-sections for $\pi^- p \to \eta n$ from 718 to 1050 mev/c,
Nucl. Phys. B \textbf{11}, 495-504 (1969).

\bibitem{Richards:1970cy}
W.~B.~Richards, C.~B.~Chiu, R.~D.~Eandi, A.~C.~Helmholz, R.~W.~Kenney, B.~Moyer, J.~A.~Poirier, R.~J.~Cence, V.~Z.~Peterson and N.~K.~Sehgal, \textit{et al.}
Production and neutral decay of the eta meson in pi- p collisions,
Phys. Rev. D \textbf{1}, 10-19 (1970).

\bibitem{Brown:1979ii}
R.~M.~Brown, A.~G.~Clark, P.~J.~Duke, W.~M.~Evans, R.~J.~Gray, E.~S.~Groves, R.~J.~Ott, H.~R.~Renshall, T.~P.~Shah and A.~J.~Shave, \textit{et al.}
Differential cross sections for the reaction $pi^-p \to \eta n$ between 724 MeV and 2723 MeV,
Nucl. Phys. B \textbf{153}, 89-111 (1979).

\bibitem{Crouch:1980vw}
H.~R.~Crouch, R.~Hargraves, R.~E.~Lanou, J.~T.~Massimo, A.~E.~Pifer, A.~M.~Shapiro, A.~E.~Brenner, M.~Ioffredo, F.~D.~Rudnick and G.~Calvelli, \textit{et al.}
Cross sections for $pi^-p \to \eta n$ and $\pi^- p \to n \eta~(\eta \to 2\gamma)$ for indident pion momenta between 1.3 and 3.8 GeV,
Phys. Rev. D \textbf{21}, 3023-3058 (1980).

\bibitem{Feltesse:1975nz}
J.~Feltesse, R.~Ayed, P.~Bareyre, P.~Borgeaud, M.~David, J.~Ernwein, Y.~Lemoigne and G.~Villet,
The Reaction $\pi^- p \to \eta n$ Up to $\rho^\ast$ ($\eta$) = 450 MeV: Experimental Results and Partial Wave Analysis,
Nucl. Phys. B \textbf{93}, 242-260 (1975).

\bibitem{Discussion:Lyu}
A private discussion with Zi-Yi Wang and professor Xiao-Rui Lyu.

\bibitem{Schmid:1967ojm}
C.~Schmid,
Final-State Interactions and the Simulation of Resonances,
Phys.\ Rev.\  {\bf 154}, no. 5, 1363 (1967).

\bibitem{Debastiani:2018xoi}
V.~R.~Debastiani, S.~Sakai and E.~Oset,
Considerations on the Schmid theorem for triangle singularities,
Eur. Phys. J. C \textbf{79}, no.1, 69 (2019).

\bibitem{Alexander:2010vd}
J.~P.~Alexander \textit{et al.} [CLEO],
Study of $\psi(2S)$ Decays to $\gamma p \bar{p}$, $\pi^0 p \bar{p}$ and $\eta p \bar{p}$ and Search for $p \bar{p}$ Threshold Enhancements,
Phys. Rev. D \textbf{82}, 092002 (2010).

\bibitem{Ablikim:2013vtm}
M.~Ablikim \textit{et al.} [BESIII],
Partial wave analysis of $\psi(2S) \to p \bar{p} \eta$,
Phys. Rev. D \textbf{88}, no.3, 032010 (2013).

\bibitem{Ablikim:2012zk}
M.~Ablikim \textit{et al.} [BESIII],
Phys. Rev. Lett. \textbf{110}, no.2, 022001 (2013)
doi:10.1103/PhysRevLett.110.022001
[arXiv:1207.0223 [hep-ex]].

\bibitem{Ablikim:2019hff}
M.~Ablikim \textit{et al.} [BESIII],
Future Physics Programme of BESIII,
Chin. Phys. C \textbf{44}, no.4, 040001 (2020).



\end{thebibliography}
\end{document}